\documentclass[twocolumn,amsmath,amssymb,superscriptaddress,reprint]{revtex4-1}
\usepackage{graphicx}
\usepackage[dvipdfmx]{color}
\usepackage{dcolumn}
\usepackage{bm}
\usepackage{txfonts}
\usepackage[utf8]{inputenc}
\usepackage[T1]{fontenc}
\usepackage{mathptmx}
\usepackage{etoolbox}
\usepackage[hypertex,colorlinks=true,linkcolor=black,citecolor=blue,filecolor=blue,urlcolor=blue,setpagesize=false,nesting=true]{hyperref}

\makeatletter
\def\@email#1#2{%
 \endgroup
 \patchcmd{\titleblock@produce}
  {\frontmatter@RRAPformat}
  {\frontmatter@RRAPformat{\produce@RRAP{*#1\href{mailto:#2}{#2}}}\frontmatter@RRAPformat}
  {}{}
}%
\makeatother
\newcommand{\msr}{$\mu$SR}
\newcommand{\sio}{Sr$_2$IrO$_4$}

\begin{document} 

\title{Direct observation of oxygen polarization in Sr$_2$IrO$_4$ by O $K$-edge x-ray magnetic circular dichroism} 

\author{R. Kadono}\thanks{ryosuke.kadono@kek.jp}
\affiliation{Muon Science Laboratory, Institute of Materials Structure Science, High Energy Accelerator Research Organization (KEK), Tsukuba, Ibaraki 305-0801, Japan}
\author{M. Miyazaki}
\affiliation{Muroran Institute of Technology, Muroran, Hokkaido 050-8585, Japan}
\author{M. Hiraishi}
\affiliation{Graduate School of Science and Engineering, Ibaraki University Bunkyo, Mito, Ibaraki, 310-8512, Japan}
\author{H. Okabe}
\affiliation{Institute for Materials Research, Tohoku University (IMR), Katahira, Aoba-ku, Sendai 980-8577, Japan}
\author{A. Koda}
\affiliation{Muon Science Laboratory, Institute of Materials Structure Science, High Energy Accelerator Research Organization (KEK), Tsukuba, Ibaraki 305-0801, Japan}
\affiliation{Department of Materials Structure Science, The Graduate University for Advanced Studies, Japan}
\author{K. Amemiya}
\affiliation{Department of Materials Structure Science, The Graduate University for Advanced Studies, Japan}
\affiliation{Photon Factory, Institute of Materials Structure Science, High Energy Accelerator Research Organization (KEK), Tsukuba, Ibaraki 305-0801, Japan}
\author{H. Nakao}
\affiliation{Department of Materials Structure Science, The Graduate University for Advanced Studies, Japan}
\affiliation{Photon Factory, Institute of Materials Structure Science, High Energy Accelerator Research Organization (KEK), Tsukuba, Ibaraki 305-0801, Japan}

\begin{abstract}
X-ray absorption spectroscopy (XAS) and magnetic circular dichroism (XMCD) measurements at the oxygen (O) $K$-edge were performed to investigate the magnetic polarization of ligand O atoms in the weak ferromagnetic (WFM) phase of the Ir perovskite compound Sr$_2$IrO$_4$. With the onset of the WFM phase below $T_{\rm N}\simeq240$ K, XMCD signals corresponding to XAS peaks respectively identified as originating from the magnetic moments of apical and planar oxygen (O$_{\rm A}$ and O$_{\rm P}$) in the IrO$_6$ octahedra were observed. The observation of magnetic moments at O$_{\rm A}$ sites is consistent  (except for the relative orientation) with that suggested by prior muon spin rotation ($\mu$SR) experiment in the non-collinear antiferromagnetic (NC-AFM) phase below $T_{\rm M}\approx100$ K.
Assuming that the O$_{\rm A}$ magnetic moment observed by $\mu$SR is also responsible for the corresponding XMCD signal, the magnetic moment of O$_{\rm P}$ is estimated to be consistent with the previous $\mu$SR result. Since the O$_{\rm A}$ XMCD signal is mainly contributed by Ir 5$d$ $zx$ and $yz$ orbitals which also hybridize with O$_{\rm P}$, it is inferred that the relatively large O$_{\rm P}$ magnetic moment is induced by Ir 5$d$ $xy$ orbitals.  Moreover, the inversion of O$_{\rm A}$ moments relative to Ir moments between the two magnetic phases revealed by XMCD suggests the presence of competing magnetic interactions for O$_{\rm A}$, with which the ordering of O$_{\rm A}$ moments in the NC-AFM phase may be suppressed to $T_{\rm M}$.
\end{abstract}
\maketitle 

Layered iridium perovskites (Sr$_{n+1}$Ir$_n$O$_{3n+1}$, $n=1, 2,$...) are interesting materials to study the unconventional properties of 5$d$ electrons arising from the competition between spin-orbit (SO) interactions, crystal fields, and Coulomb interactions (electron correlations) with comparable energies. Recent studies using various microscopic probes have revealed that the monolayer compound \sio\ can be regarded as a Mott insulator realized in the $t_{2g}$ multiplet reorganized by the relatively strong SO interaction. The 5$d^5$ (Ir$^{4+}$) electrons occupy the $t_{2g}$ states with effective angular momentum $L_{\rm eff} = 1$ and spin $S=1/2$ are split into $J_{\rm eff} =1/2$ doublet and $J_{\rm eff} =3/2$ quartet, in which a gap is induced in the doublet band by the Coulomb repulsion, resulting in an insulator state with pseudo-spin $J_{\rm eff} =1/2$ that consists of three spin-orbit components in the $t_{2g}$ band with equal weights: 
$$|J_{\rm eff} =1/2,\pm1/2\rangle = \frac{1}{\sqrt{3}}(|xy,\pm\sigma\rangle\pm|yz,\mp\sigma\rangle+i|zx,\mp\sigma\rangle),$$ 
where $\sigma$ denotes the spin state \cite{Kim:08,Kim:09,Jackeli:09,Wang:11}. While the $J_{\rm eff} =1/2$ model is supported by resonant and inelastic x-ray scattering \cite{Kim:09,Kim:12,Bertinshaw:19}, the simple description with equal weights for the orbital populations has been called into question due to a non-negligible tetragonal distortion of the IrO$_6$ octahedra \cite{Chapon:11,Haskel:12,Moretti:14,Jeong:20,Perkins:14}.

\begin{figure}[h]
	\begin{center}
\includegraphics[width=0.85\linewidth]{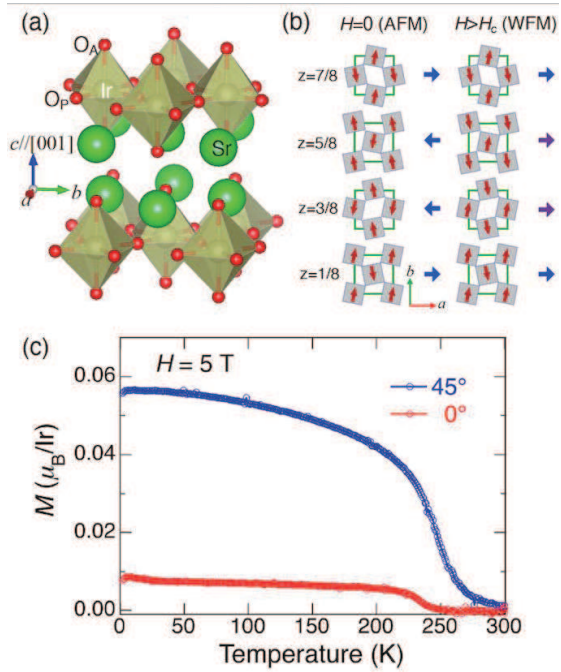}
			\caption{(a) Crystal structure of \sio\ viewed from the $a$ axis, where O$_{\rm P}$ and O$_{\rm A}$ refer to planar and apical oxygen (drawn by VESTA \cite{Vesta}).  (b) Non-collinear AFM ordering pattern of $J_{\rm eff} = 1/2$ moments (arrows) within IrO$_2$ planes and their stacking pattern along the $c$ axis in zero field and in the WFM state (after Ref.\cite{Kim:09}). (c) Uniform magnetization ($M$) vs.~temperature in \sio\ at $H=5$ T  with the polar angle between the $c$ axis and $H$ being 0$^\circ$ and 45$^\circ$. }
			\label{mag}
	\end{center}
\end{figure}

\sio\ crystalizes in a tetragonal structure with $I4_1/acd$ space group [see Fig.~\ref{mag}(a)], and exhibits antiferromagnetic (AFM) order at the transition temperature $T_N\simeq 240$ K under zero external field \cite{Crawford:94,Huang:94,Kim:09,Ge:11,Ye:13,Dhital:13,Sung:16}.  The AFM phase exhibits an extra feature of non-collinear (NC) Ir spin canting ($\simeq$11$^\circ$) within the $ab$ plane [Fig.~\ref{mag}(b), left], which is mainly attributed to the rotation of IrO$_6$ octahedra by the lattice distortion \cite{Kim:08}.  It also undergoes metamagnetic transition to weak ferromagnetic (WFM) phase due to the alignment of Ir canting along the $a$ axis above $H_c\approx 0.2$ T [Fig.~\ref{mag}(b), right]. 

Nevertheless, that the NC-AFM phase does not necessarily fit within the framework of the $J_{\rm eff} =1/2$ model is also evident in its complex magnetic properties at lower temperatures. It was shown that bulk magnetization [$M(H)$] decreases anomalously at $T<T_{\rm M} \lesssim 100$ K and $H<H_c$, suggesting that AFM correlations are enhanced below $T_{\rm M}$ \cite{Chikara:09,Li:13}. The decrease in the Ir-O-Ir bond angle in correlation with $M$ suggests that this is related with a change in the exchange coupling \cite{Li:13}. Moreover, anomalous behavior of frequency-dependent ac-dielectric loss is reported around $T_{\rm M}$ \cite{Chikara:09}, strongly suggesting the critical slowing down of electric polarization fluctuation \cite{Katsura:05,Nagaosa:08}. Therefore, it is natural to infer the existence of electro-magnetic cross-correlations in the NC-AFM due to the inverse Dzyaloshinsky-Moriya (DM) mechanism (or the ``spin current" mechanism \cite{Sergienko:06}). While the DM interaction seems to have little role in the the $J_{\rm eff} =1/2$ framework, the rotation of the IrO$_6$ octahedra generates a DM and a $J_z$ term to the magnetic interaction, which limits the mutual angle between adjacent pseudospins to $\pi+2\alpha$ with the octahedron rotation angle $\alpha$  \cite{Jackeli:09}. 

Our previous muon spin rotation (\msr) measurements of the internal magnetic field ($B_{\rm loc}$) in \sio\ revealed a gradual development of the second stage of magnetic order below $T_{\rm M}$ \cite{Miyazaki:15}. Furthermore, the $B_{\rm loc}$ felt by muons occupying sites close to the apical oxygen (O$_{\rm A}$) of the IrO$_6$ octahedra showed a further increase, which can only be consistent with the selective appearance of ordered magnetic moments ($\mu_{\rm O_A}\approx$0.03$\mu_B$ at $T\rightarrow0$, antiparallel with the Ir moments) at the O$_{\rm A}$ sites.  This is reminiscent of the correlation between charge transfer to O and associated orbital polarization and the non-collinear Mn spin structure in the multiferroic  YMn$_2$O$_5$ that we recently found by \msr\ and resonant x-ray scattering, supporting the multiferroic scenario for the charge sector anomaly reported in the low $T$-$H$ region \cite{Ishii:20}. 

However, it is difficult to determine only by \msr\ as a local probe whether the O moments are in a long-range ordered state.  Specifically, the possibility remains that the muon, as a pseudo-hydrogen, forms a hydroxyl-like OMu state that accompanies an electron localized on the adjacent magnetic ion (i.e., a polaron state) to modulate $B_{\rm loc}$, as has recently been pointed out for other magnetic oxides \cite{Dehn:20,Dehn:21}.  
In any case, there is still no other information of O polarization in \sio, remaining at the circumstantial stage.

 The x-ray absorption spectroscopy (XAS) and magnetic circular dichroism (XMCD) at the O $K$-absorption edge can provide useful information on O holes (unoccupied 2$p$ orbitals) and their magnetic polarization in transition metal oxides, because the 2$p$ orbitals are strongly linked with the ground state properties; it was used to reveal the mechanism of the metal-insulator transition of La$_{1-x}$Sr$_x$MnO$_3$  \cite{Koide:01} and the evolution of O orbital moment order in La$_{1-x}$Sr$_x$CoO$_3$ \cite{Okamoto:00,Medling:12}. Furthermore, the O $K$-edge XMCD of CrO$_2$ has been used to explicitly demonstrate the delocalization of the hybridized states of Cr 3$d$ and O 2$p$ \cite{Goering:02}. 
 
 Here, we present a study using O $K$-edge XAS and XMCD in the WFM phase of \sio\ to examine the possible O hole polarization due to O 2$p$-Ir 5$d$ orbital hybridization.   The XMCD signal was observed at 5 T in accordance with the development of the WFM phase below $T_N$, where the signal from the planar oxygen (O$_{\rm P}$) is significantly enhanced compared to that from O$_{\rm A}$.  The temperature dependence of the XMCD signal (with negative sign) is nearly proportional to the bulk magnetization $M(H)$ measured under similar conditions, suggesting that magnetic moments of both O$_{\rm P}$ and O$_{\rm A}$ are parallel to those of Ir and their magnitude are proportional. Assuming that the O$_{\rm A}$ magnetic moment observed by $\mu$SR in the NC-AFM phase is also responsible for the corresponding XMCD signal, the magnetic moment of O$_{\rm P}$ is extrapolated to be $\mu_{\rm O_P}\approx0.10(1)\mu_B$ from the ratio of XMCD signal intensity between O$_{\rm P}$ and O$_{\rm A}$.  Since the O$_{\rm A}$ signal is mainly contributed by Ir 5$d$ $zx$ and $yz$ orbitals which also hybridize with O$_{\rm P}$, the relatively large $\mu_{\rm O_P}$ is mainly attributed to the Ir 5$d$ $xy$ orbitals.  This is consistent with recent reports that the $xy$ orbitals predominantly contribute to the magnetism of Ir \cite{Jeong:20}. 
Notably, the orientation of the O$_{\rm A}$ moments relative to Ir moments in the WFM phase is reversed from that in the NC-AFM phase, suggesting that competing magnetic correlations are acting on O$_{\rm A}$, leading to the suppression of O$_{\rm A}$ ordering to $T_{\rm M}$ in the NC-AFM phase.

\begin{figure}[t]
	\begin{center}
\includegraphics[width=0.8\linewidth]{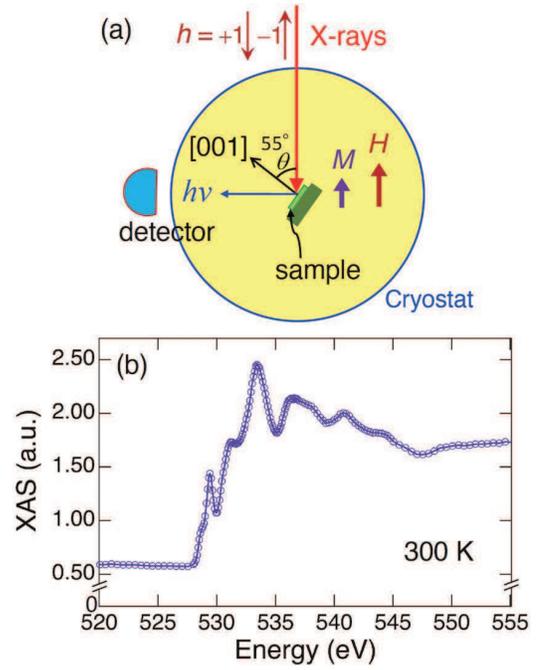}
			\caption{(a) Experimental conditions for XMCD measurement: a circular polarized x-ray beam (helicity $h=\pm1$ defined by arrows) was incident at $\theta=55^\circ$ to the $c$ axis of the \sio\ single crystal, and the magnetic field ($H$) was parallel to the x-ray. The fluorescence x-ray was measured by a solid-state detector placed perpendicular to the x-ray beam direction. (b) Example of absorption spectrum (raw data) observed at 300 K.}
			\label{xas}
	\end{center}
\end{figure} 

Single-crystal samples were grown by the self-flux method in platinum crucibles with Ir, SrCO$_3$, and SrCl$_2$ weighed and mixed in molar ratios of 1:2:7, respectively. The sintering conditions were from 1300 $^\circ$C to 900 $^\circ$C at a rate of 8 $^\circ$C/hr, followed by cooling down to room temperature. Single crystals of approximately 1 mm$^2\times3$ mm (7--9 mg) in size were obtained, which were flaky and layered with a metallic luster. X-ray diffraction of the flake surface obtained with adhesive tape showed a $c$ plane of \sio, and no Sr$_3$Ir$_2$O$_7$ phase (which is easily stacked as an impurity) was detected.  $M(H)$ measurements with $H$ parallel and perpendicular to the $c$ axis showed a clear AFM/WFM transition around 240--250 K. At a low field ($H=0.1$ T, $\perp c$), $M$ showed a relatively rapid increase from 240 K to 220 K, followed by a gradual concave decrease with decreasing temperature. These behaviors of $M(H)$ indicates that the samples were of the same high quality as reported in previous studies. The temperature dependence of $M$ measured by MPMS (Quantum Design) under $H=5$ T (corresponding to the WFM phase) with the polar angle between the $c$ axis and $H$ being 0$^\circ$ and 45$^\circ$ is shown in Fig.~\ref{mag}(c). 

The XAS-XMCD measurements were performed using circularly polarized synchrotron radiation from the undulator beamline BL-16A at the Photon Factory, High Energy Accelerator Research Organization (KEK) \cite{Amemiya:10}.  A single crystal with $c$ surface obtained by exfoliation was mounted on the copper cold finger with silver grease, and sample temperature was controlled by a He gas-flow cryostat. The XAS spectra were obtained in the partial-fluorescence-yield mode over the energy range of 520--555 eV, centered around the O $K$-edge ($\sim$530 eV) with alternative polarization (helicity $h= \pm1$, where positive sign corresponds to the direction opposite to $M$), and temperature was scanned from 10 to 300 K. Due to geometrical restrictions among incident x-rays, superconducting magnet, and the x-ray detector, the experiment was conducted in the arrangement schematically shown in Fig.~\ref{xas}(a): the incident x-ray beam was parallel with the external magnetic field, and it was obliquely incident at $\theta= 55^\circ$ (``magic angle'' for eliminating the contribution of magnetic dipolar term) to the $c$ axis of the sample crystal.  Since BL-16A is capable of polarization switching at about 10 Hz in combination with a kicker magnet \cite{Amemiya:13}, an attempt was made to detect small changes in magnetic-field-induced O polarization by using the differential spectroscopy to suppress the effect of drift over a few hours. 
In addition, to eliminate systematic errors due to the different beam properties of the two undulators, XMCD measurements were made on two alternative combinations of mutual undulator polarity.  Fig.~\ref{xas}(b) shows an example of raw XAS data observed at 300 K. 

\begin{figure}[t]
	\begin{center}
\includegraphics[width=0.9\linewidth]{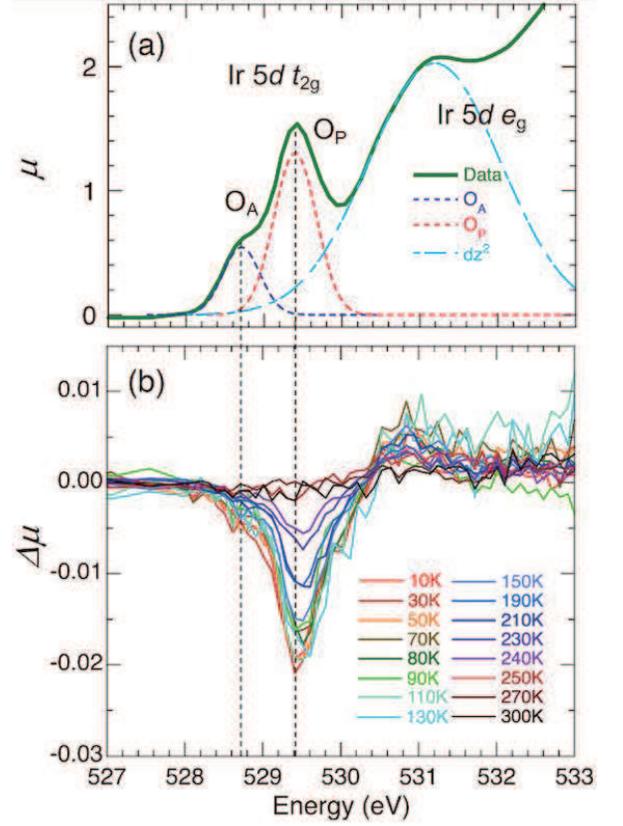}
			\caption{(a) normalized XAS spectrum observed at 300 K in \sio, where dashed lines represent the result of curve fit assuming three Gaussian peaks yielding 528.70(4), 529.40(2), and 531.19(5) eV for the peak energy; O$_{\rm A}$ and O$_{\rm P}$ indicate contributions from the apical and planar O hybridized with Ir $5d$ $t_{2g}$ orbitals, whereas $d_{z^2}$ shows that from O$_{\rm A}$ hybridized with $e_{g}$ orbitals. (b) XMCD spectra [$\varDelta\mu(E)$] at all temperatures measured in \sio.}
			\label{xasmcd}
	\end{center}
\end{figure}

The helicity-dependent XAS spectra, $\mu_\pm(E)=I_\pm(E)/I_0$, were obtained by subtracting averaged offset for $520\le E\le 525$ eV as background and normalized by the average intensity $I_0$ for $I_\pm(E>550\:{\rm eV})$ [see Fig.~\ref{xas}(b)]. It turned out that the total XAS spectrum [$\mu(E)=\mu_+(E)+\mu_-(E)$] was independent of temperature within the range of experimental accuracy. An example of $\mu(E)$ obtained at 300 K are shown in Fig.~\ref{xasmcd}(a).  The peaks around 528--530 eV and 530--535 eV reproduce the previously reported O $K$-edge XAS spectra which are respectively assigned to the $t_{2g}$ and $e_g$ bands of Ir 5$d$ mixed with O 2$p$ \cite{Kim:08,Moon:06,Liu:15,Ilakovac:19}. This is in good agreement with the estimation of the lowest energy unoccupied orbitals to be in the Ir $5d$ band from the partial density of states obtained from band calculations \cite{Bhandari:19}, and we can conclude that the spectra in Fig.~\ref{xasmcd}(a) are assigned in the same manner. Furthermore, from comparison with the linear-polarization dependence of XAS reported earlier, we can infer that the peaks at 528.70(4) eV and  529.40(2) eV are respectively contributed by O$_{\rm A}$ and  O$_{\rm P}$ of the IrO$_6$ octahedra \cite{Moon:06,Ilakovac:19,Lu:18}.  The relative XAS signal intensity of O$_{\rm A}$ and O$_{\rm P}$ in the 300 K data is estimated to be $1:2.39(24)$ from the curve fit [see Fig.~\ref{xasmcd}(a)].

Here, let us examine the influence of polarization to the relative XAS intensity between O$_{\rm A}$ and O$_{\rm P}$.  For circularly polarized light, the $\theta$ dependence of $\mu$ is given by  $\frac{1}{2}(\cos^2\theta + 1)n_{yz/zx}$ for O$_{\rm A}$ and $\frac{1}{2}(\cos^2\theta + 1)n_{xy}+\frac{1}{2}(\sin^2\theta)n_{yz/zx}$ for O$_{\rm P}$, where $n_{xy}$ and $n_{yz/zx}$ are the number of holes in the $xy$ and $yz/zx$ states hybridized with $2p$ orbitals \cite{Mizokawa:01}.  The ratio $n_{xy}/n_{yz/zx}\equiv R$ is then deduced from the equation 
\begin{equation}
\frac{\frac{1}{2}(\cos^2\theta + 1)R+\frac{1}{2}\sin^2\theta}{\frac{1}{2}(\cos^2\theta + 1)}=2.39(24),
\end{equation}
which yields $R=1.89(19)$ for the present case of $\theta=55^\circ$. This value suggests that the relative hole occupancy for $xy$, $yz$, and $zx$ orbitals is approximately $1:1:1$.

\begin{figure}[t]
	\begin{center}
\includegraphics[width=0.9\linewidth]{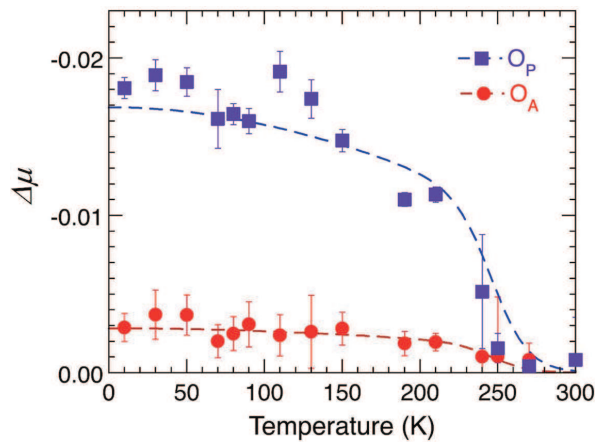}
			\caption{Temperature dependence of XMCD signals at 528.8 eV (O$_{\rm A}$) and 529.5 eV (O$_{\rm P}$) in \sio. Dashed lines indicate the results of curve fits by scaling $M(T)$ at $H=5$ T oriented 45$^\circ$ from $c$ axis [see Fig.~\ref{mag}(c)].}
			\label{xmcd}
	\end{center}
\end{figure}

The XMCD spectrum is defined as
\begin{equation}
\varDelta\mu(E)= \mu_+(E) - \mu_-(E),
\end{equation}
and the spectra for each measurement temperature are shown in Fig.~\ref{xasmcd}(b), where the O$_{\rm P}$ signal clearly increases with decreasing temperature from around $T_N$. A similar tendency is also observed for the O$_{\rm A}$ signal. These two peaks and the structure around 531 eV are approximated by three Gaussian distributions, and the peak intensities are estimated by curve fitting for $\varDelta\mu(527\le E\le 532$ eV). (The positive peak at 531 eV is considered an artifact because the intensity is only $10^{-3}$ of the corresponding XAS signal.) As shown in Fig.~\ref{xmcd}, the temperature dependence of the peak intensity roughly follows that of $M$ measured under the similar conditions [$H=5$ T, 45$^\circ$, see Fig.~\ref{mag}(c)]: the dashed curves in Fig.~\ref{xmcd} are obtained by curve fitting using a form
$$\varDelta\mu=\varDelta\mu_0\frac{M(T)}{M(0)}$$
with only the normalization factor $\varDelta\mu_0$ as a parameter [yielding $\varDelta\mu_0= -0.0169(3)$ and $-0.0028(3)$ for O$_{\rm P}$ and O$_{\rm A}$, respectively].  This indicates that the O 2$p$ magnetic moments are polarized, confirming that the polarization of the O 2$p$ state is induced by hybridization between the O 2$p$ and Ir 5$d$ orbitals.
 
 The negative sign of the O$_{\rm A/P}$ signals implies that the orientation of the O 2$p$ orbital magnetic moment is in the same direction as that of $H$ (and $M$). The value of $I_0$ used in deducing $\mu_\pm$ was roughly the same as the value of $I_\pm$ for the O$_{\rm P}$ signal [see Fig.~\ref{xas}(b)], and hardly changed with temperature. Thus, the value of $\varDelta\mu$ appears to suggest that the mixing between O 2$p$ and Ir 5$d$ is at most a few \%. (Because the ambiguity in the energy range of signal integral makes it difficult to apply the sum rules \cite{Thole:92,Carra:93} to the XMCD spectra at the O $K$-edge, we did not estimate the magnitude of the orbital moment from its signal intensity.)  Moreover, it should be noted that XMCD is proportional to the net average moment which is biased by the residual ferromagnetic component associated with the WFM phase for $H>H_c$ ($\approx0.08\mu_B$/Ir plus O moment contribution)].  
 
 However, the small $\varDelta\mu$ does not necessarily mean that orbital hybridization is small, given the signal intensity originally expected for O $K$-edge XMCD; the SO interaction to split O 2$p_{1/2}$ and 2$p_{3/2}$ levels is generally considered to be weak due to the small gradient of the Coulomb potential.  Indeed, a strong hybridization of the O 2$p$ and Ir 5$d$ orbitals has been assumed to explain the significant reduction of the ordered Ir magnetic moment ($\mu_{\rm Ir}\approx 0.4\mu_B$) \cite{Kim:08}, the AFM exchange interaction between neighboring Ir ions, and the tilted Ir moment associated with octahedral rotation \cite{Jackeli:09,Perkins:14}.  Therefore, the results in Fig.~\ref{xmcd} are only to suggest that the orientations and magnitudes of the O orbital moments relative those of Ir are nearly unchanged regardless of temperature in the WFM phase.

It is noticeable in Figs.~\ref{xasmcd} and \ref{xmcd} that, in view of the relative XAS signal intensity between O$_{\rm A}$ and O$_{\rm P}$, the XMCD signal from O$_{\rm P}$ is significantly enhanced compared to that from O$_{\rm A}$ [$= 1:6.04(11)$].  To theoretically explore its origin, we calculated XAS spectra and partial density of states using the FDMNES code \cite{Bunau:09,Guda:15}, and found that the Ir $d_{xy}$ orbital is exclusively coupled with O$_{\rm P}$ (see Supplemental Material \cite{SM} for details).  Assuming that the origin of the O$_{\rm A}$ XMCD signal is common to  $\mu_{\rm O_A}$ inferred from $\mu$SR in the NC-AFM phase, the O$_{\rm P}$ moment size can be evaluated by scaling that of  $\mu_{\rm O_A}$ ($\approx0.03\mu_B$, $T\rightarrow0$) by the ratio of the XMCD signals of O$_{\rm P}$ and O$_{\rm  A}$ normalized by the XAS intensity, $6.04(11)/2 = 3.02(4)$, yielding $\mu_{\rm O_P}=0.10(1)\mu_B$. 

 In the ZF-$\mu$SR measurement, it is difficult to distinguish the O$_{\rm P}$ contribution proportional to Ir magnetization from that of Ir, and the magnetic polarization of O$_{\rm P}$ was not considered in the previous $\mu$SR study \cite{Miyazaki:15}. Therefore, we reexamined whether the occurrence of the magnetic polarization of O$_{\rm P}$ is consistent with the previous $\mu$SR result. Specifically, assuming $\mu_{\rm Ir}\approx0.4\mu_B$ and that both O$_{\rm A}$ and O$_{\rm P}$ have the same ratio of spin to orbital moments and that their magnitudes do not change with the magnetic field, we simulated the internal magnetic field at the muon sites by changing the size of $\mu_{\rm O_P}$. As a result, it turned out that the simulation reproduces the observed $\mu$SR spectra when $\mu_{\rm O_P}\approx0.10\mu_B$ in the same direction as Ir, which is in excellent agreement with that extrapolated from O$_{\rm P}$ using the XAS-XMCD result. Thus, assuming that $\mu_{\rm O_A}$ observed by $\mu$SR in the NC-AFM phase is also responsible for the corresponding XMCD signal in the WFM phase, the XMCD and $\mu$SR results are found to be perfectly in line with each other (except for the $\mu_{\rm O_A}$ orientation relative to that of $\mu_{\rm Ir}$). That both $\mu_{\rm O_P}$ and $\mu_{\rm O_A}$ are small is also consistent with the absence of a corresponding signal in the polarized neutron experiment in the WFM phase \cite{Jeong:20}.

The earlier XAS measurements using linear-polarized light indicate that the O$_{\rm A}$ signal mainly reflects the contribution of unoccupied 2$p_x$/2$p_y$ orbitals produced by the hybridization with the Ir 5$d$ $zx$/$yz$ orbitals, while  the O$_{\rm P}$ involves all of the 5$d$ orbitals \cite{Lu:18}; FDMNES calculations suggest that this is also the case for circularly polarized light (see SM \cite{SM}). From the partial density of states for O$_{\rm P}$ XAS signal obtained from the FDMNES calculation, the relative contribution of the $xy$ orbital to those of $zx$ and $yz$ is evaluated to be $1 : 1.9$, in good agreement with the experimental evaluation.  Thus, the relatively large O$_{\rm P}$ signal suggests that the Ir 5$d$ $xy$ orbitals are mainly responsible for the O$_{\rm P}$ polarization.  This is consistent with recent reports that the $xy$ orbitals are dominant in the Ir 5$d$ ground state \cite{Chapon:11,Haskel:12,Moretti:14,Jeong:20}.

The relationship between the temperature dependence of the XMCD signal in the WFM phase ($H>H_c$) and that of the spontaneous order of O$_{\rm A}$ magnetic moments suggested by $\mu$SR ($H=0$) is not obvious at this stage.  The latter is observed only in the NC-AFM phase below around $T_{\rm M}$ and reaches a maximum moment size of $\mu_{\rm O_A}\approx0.03\mu_B$ below $\sim$20 K. Nevertheless, considering that the direction of the O$_{\rm A}$ moments in the NC-AFM phase below $T_{\rm M}$ is reversed from that in the WFM phase, it is strongly suggested that competing magnetic correlations are acting on O$_{\rm A}$ in the NC-AFM phase,  suppressing the ordering of O$_{\rm A}$ moments for $T>T_{\rm M}$. In the WFM phase, on the other hand, the AFM-like correlations are suppressed by the external magnetic field, and the O$_{\rm A}$ moments are interpreted as aligned with the Ir magnetic moments over the whole temperature range below $T_{\rm N}$. 

 Concerning the anomalous ac-dielectric loss below $T_{\rm M}$ in the NC-AFM phase \cite{Chikara:09}, we point out that the Ir spins can be regarded to take on a cycloidal magnetic structure when the magnetic modulation vectors (${\bm V}_{ij}$) are chosen in some specific directions connecting the IrO$_2$ planes (see Supplemental Material \cite{SM} including Refs.\cite{Kenzelmann:05,Yamazaki:07,Bogdanov:15}). This suggests that electric polarization (${\bm p}$) perpendicular to ${\bm V}_{ij}$ can be induced by the inverse DM mechanism.
Such a cycloidal magnetic structure is always paired with one showing inverted spin rotation generating an opposite electric polarization ($-{\bm p}$), and therefore considered to induce an anti-ferroelectric correlation without producing macroscopic polarization.
Interestingly, it has been reported that the ac-dielectric loss in the $a$-axis direction below $T_{\rm M}$ is nearly twice as large as that in the $c$-axis direction \cite{Chikara:09}. This anisotropic response is consistent with what would be expected from the aforementioned direction of electric polarization and supports the scenario due to the inverse DM mechanism.

It should be noted that the temperature dependence of XMCD intensity in Fig.~\ref{xmcd} seems to deviate considerably  from that of $M$ at lower temperatures: it peaks around 40 K and 120 K and shows a minimum around $T_{\rm M}$. This trend is also confirmed for the integrated signals of the two peaks ($527.9\le E\le 530.5$ eV, not shown) in the XMCD spectra, suggesting that the orbital hybridization may vary slightly with temperature. It is tempting to imagine that this variation is also related to the anomalies described above \cite{SM}, but further study will be needed to determine whether this is the case.

Finally, we note that the observed O $K$-edge XMCD is unusually large compared to that in the $3d$ compounds. Since the Ir $5d$ electrons are subject to the strong SO interaction and their magnetization is related to a stronger orbital contribution than in most of the $3d$ transition metals, it is interesting to compare the relative O $K$-edge XMCD intensity to the magnitude of the average Ir magnetic moment in this context. Specifically, the relative XMCD intensity of the O$_{\rm P}$ peak ($\varDelta\mu=\varDelta\mu_0\simeq 0.017$ divided by $\mu\simeq1.55$) is compared to the projected moment of Ir ($\overline{\mu}_{\rm Ir}\simeq0.056\mu_B$) to yield $(\varDelta\mu/\mu)/\overline{\mu}_{\rm Ir}\simeq0.20\mu_B^{-1}$. Comparing this to the ferromagnetic CrO$_2$ along the $c$-axis \cite{Goering:02}, the similar estimation yields $(\varDelta\mu/\mu)/\overline{\mu}_{\rm Cr}\simeq0.017\mu_B^{-1}$ which is $\sim$14 times smaller than that for Ir. Thus, the O $K$-edge XMCD induced by Ir magnetic moment is more than an order of magnitude greater than that of the normal $3d$ ferromagnets. 

In summary, we have demonstrated by O $K$-edge XAS-XMCD measurements that ligand O has finite orbital magnetic moments induced by hybridization with Ir 5$d$ orbitals in the WFM phase of \sio.  We combined the present result with that of the previous $\mu$SR in the NC-AFM phase to evaluate the magnitude of the weak magnetic moments at the O$_{\rm P}$ and O$_{\rm A}$ sites.  The orientation of the O$_{\rm A}$ moments relative to Ir moments inferred by XMCD is reversed from that in the NC-AFM phase, suggesting that competing magnetic correlations are acting on O$_{\rm A}$, leading to the suppression of O$_{\rm A}$ ordering to $T_{\rm M}$ in the latter phase. The magnetic polarization of O holes confirms modulation in charge distribution involving ligand O, and it also suggests that anti-ferroelectric correlations due to the inverse Dzyaloshinsky-Moriya mechanism is likely responsible for the observed anomalous dielectric loss in the NC-AFM phase.

We thank Y. Joly for help in using the FDMNES code. Thanks are also to H. Sagayama for helpful discussion. The XAS/XMCD experiments were conducted under the support of Inter-University-Research Programs by Institute of Materials Structure Science, KEK (Proposal No.~2021G544). This work was partially supported by JSPS KAKENHI (Grant No.~19K15033).  We also acknowledge the Neutron Science and Technology Center, CROSS for the use of PPMS in their user laboratories.

\begin{thebibliography}{47}%
\makeatletter
\providecommand \@ifxundefined [1]{%
 \@ifx{#1\undefined}
}%
\providecommand \@ifnum [1]{%
 \ifnum #1\expandafter \@firstoftwo
 \else \expandafter \@secondoftwo
 \fi
}%
\providecommand \@ifx [1]{%
 \ifx #1\expandafter \@firstoftwo
 \else \expandafter \@secondoftwo
 \fi
}%
\providecommand \natexlab [1]{#1}%
\providecommand \enquote  [1]{``#1''}%
\providecommand \bibnamefont  [1]{#1}%
\providecommand \bibfnamefont [1]{#1}%
\providecommand \citenamefont [1]{#1}%
\providecommand \href@noop [0]{\@secondoftwo}%
\providecommand \href [0]{\begingroup \@sanitize@url \@href}%
\providecommand \@href[1]{\@@startlink{#1}\@@href}%
\providecommand \@@href[1]{\endgroup#1\@@endlink}%
\providecommand \@sanitize@url [0]{\catcode `\\12\catcode `\$12\catcode
  `\&12\catcode `\#12\catcode `\^12\catcode `\_12\catcode `\%12\relax}%
\providecommand \@@startlink[1]{}%
\providecommand \@@endlink[0]{}%
\providecommand \url  [0]{\begingroup\@sanitize@url \@url }%
\providecommand \@url [1]{\endgroup\@href {#1}{\urlprefix }}%
\providecommand \urlprefix  [0]{URL }%
\providecommand \Eprint [0]{\href }%
\providecommand \doibase [0]{http://dx.doi.org/}%
\providecommand \selectlanguage [0]{\@gobble}%
\providecommand \bibinfo  [0]{\@secondoftwo}%
\providecommand \bibfield  [0]{\@secondoftwo}%
\providecommand \translation [1]{[#1]}%
\providecommand \BibitemOpen [0]{}%
\providecommand \bibitemStop [0]{}%
\providecommand \bibitemNoStop [0]{.\EOS\space}%
\providecommand \EOS [0]{\spacefactor3000\relax}%
\providecommand \BibitemShut  [1]{\csname bibitem#1\endcsname}%
\let\auto@bib@innerbib\@empty
\bibitem [{\citenamefont {Kim}\ \emph {et~al.}(2008)\citenamefont {Kim},
  \citenamefont {Jin}, \citenamefont {Moon}, \citenamefont {Kim}, \citenamefont
  {Park}, \citenamefont {Leem}, \citenamefont {Yu}, \citenamefont {Noh},
  \citenamefont {Kim}, \citenamefont {Oh}, \citenamefont {Park}, \citenamefont
  {Durairaj}, \citenamefont {Cao},\ and\ \citenamefont {Rotenberg}}]{Kim:08}%
  \BibitemOpen
  \bibfield  {author} {\bibinfo {author} {\bibfnamefont {B.~J.}\ \bibnamefont
  {Kim}}, \bibinfo {author} {\bibfnamefont {H.}~\bibnamefont {Jin}}, \bibinfo
  {author} {\bibfnamefont {S.~J.}\ \bibnamefont {Moon}}, \bibinfo {author}
  {\bibfnamefont {J.-Y.}\ \bibnamefont {Kim}}, \bibinfo {author} {\bibfnamefont
  {B.-G.}\ \bibnamefont {Park}}, \bibinfo {author} {\bibfnamefont {C.~S.}\
  \bibnamefont {Leem}}, \bibinfo {author} {\bibfnamefont {J.}~\bibnamefont
  {Yu}}, \bibinfo {author} {\bibfnamefont {T.~W.}\ \bibnamefont {Noh}},
  \bibinfo {author} {\bibfnamefont {C.}~\bibnamefont {Kim}}, \bibinfo {author}
  {\bibfnamefont {S.-J.}\ \bibnamefont {Oh}}, \bibinfo {author} {\bibfnamefont
  {J.-H.}\ \bibnamefont {Park}}, \bibinfo {author} {\bibfnamefont
  {V.}~\bibnamefont {Durairaj}}, \bibinfo {author} {\bibfnamefont
  {G.}~\bibnamefont {Cao}}, \ and\ \bibinfo {author} {\bibfnamefont
  {E.}~\bibnamefont {Rotenberg}},\ }\href {\doibase
  10.1103/PhysRevLett.101.076402} {\bibfield  {journal} {\bibinfo  {journal}
  {Phys. Rev. Lett.}\ }\textbf {\bibinfo {volume} {101}},\ \bibinfo {pages}
  {076402} (\bibinfo {year} {2008})}\BibitemShut {NoStop}%
\bibitem [{\citenamefont {Kim}\ \emph {et~al.}(2009)\citenamefont {Kim},
  \citenamefont {Ohsumi}, \citenamefont {Komesu}, \citenamefont {Sakai},
  \citenamefont {Morita}, \citenamefont {Takagi},\ and\ \citenamefont
  {Arima}}]{Kim:09}%
  \BibitemOpen
  \bibfield  {author} {\bibinfo {author} {\bibfnamefont {B.~J.}\ \bibnamefont
  {Kim}}, \bibinfo {author} {\bibfnamefont {H.}~\bibnamefont {Ohsumi}},
  \bibinfo {author} {\bibfnamefont {T.}~\bibnamefont {Komesu}}, \bibinfo
  {author} {\bibfnamefont {S.}~\bibnamefont {Sakai}}, \bibinfo {author}
  {\bibfnamefont {T.}~\bibnamefont {Morita}}, \bibinfo {author} {\bibfnamefont
  {H.}~\bibnamefont {Takagi}}, \ and\ \bibinfo {author} {\bibfnamefont
  {T.}~\bibnamefont {Arima}},\ }\href {\doibase 10.1126/science.1167106}
  {\bibfield  {journal} {\bibinfo  {journal} {Science}\ }\textbf {\bibinfo
  {volume} {323}},\ \bibinfo {pages} {1329} (\bibinfo {year}
  {2009})}\BibitemShut {NoStop}%
\bibitem [{\citenamefont {Jackeli}\ and\ \citenamefont
  {Khaliullin}(2009)}]{Jackeli:09}%
  \BibitemOpen
  \bibfield  {author} {\bibinfo {author} {\bibfnamefont {G.}~\bibnamefont
  {Jackeli}}\ and\ \bibinfo {author} {\bibfnamefont {G.}~\bibnamefont
  {Khaliullin}},\ }\href {\doibase 10.1103/PhysRevLett.102.017205} {\bibfield
  {journal} {\bibinfo  {journal} {Phys. Rev. Lett.}\ }\textbf {\bibinfo
  {volume} {102}},\ \bibinfo {pages} {017205} (\bibinfo {year}
  {2009})}\BibitemShut {NoStop}%
\bibitem [{\citenamefont {Wang}\ and\ \citenamefont {Senthil}(2011)}]{Wang:11}%
  \BibitemOpen
  \bibfield  {author} {\bibinfo {author} {\bibfnamefont {F.}~\bibnamefont
  {Wang}}\ and\ \bibinfo {author} {\bibfnamefont {T.}~\bibnamefont {Senthil}},\
  }\href {\doibase 10.1103/PhysRevLett.106.136402} {\bibfield  {journal}
  {\bibinfo  {journal} {Phys. Rev. Lett.}\ }\textbf {\bibinfo {volume} {106}},\
  \bibinfo {pages} {136402} (\bibinfo {year} {2011})}\BibitemShut {NoStop}%
\bibitem [{\citenamefont {Kim}\ \emph {et~al.}(2012)\citenamefont {Kim},
  \citenamefont {Casa}, \citenamefont {Upton}, \citenamefont {Gog},
  \citenamefont {Kim}, \citenamefont {Mitchell}, \citenamefont {van
  Veenendaal}, \citenamefont {Daghofer}, \citenamefont {van~den Brink},
  \citenamefont {Khaliullin},\ and\ \citenamefont {Kim}}]{Kim:12}%
  \BibitemOpen
  \bibfield  {author} {\bibinfo {author} {\bibfnamefont {J.}~\bibnamefont
  {Kim}}, \bibinfo {author} {\bibfnamefont {D.}~\bibnamefont {Casa}}, \bibinfo
  {author} {\bibfnamefont {M.~H.}\ \bibnamefont {Upton}}, \bibinfo {author}
  {\bibfnamefont {T.}~\bibnamefont {Gog}}, \bibinfo {author} {\bibfnamefont
  {Y.-J.}\ \bibnamefont {Kim}}, \bibinfo {author} {\bibfnamefont {J.~F.}\
  \bibnamefont {Mitchell}}, \bibinfo {author} {\bibfnamefont {M.}~\bibnamefont
  {van Veenendaal}}, \bibinfo {author} {\bibfnamefont {M.}~\bibnamefont
  {Daghofer}}, \bibinfo {author} {\bibfnamefont {J.}~\bibnamefont {van~den
  Brink}}, \bibinfo {author} {\bibfnamefont {G.}~\bibnamefont {Khaliullin}}, \
  and\ \bibinfo {author} {\bibfnamefont {B.~J.}\ \bibnamefont {Kim}},\ }\href
  {\doibase 10.1103/PhysRevLett.108.177003} {\bibfield  {journal} {\bibinfo
  {journal} {Phys. Rev. Lett.}\ }\textbf {\bibinfo {volume} {108}},\ \bibinfo
  {pages} {177003} (\bibinfo {year} {2012})}\BibitemShut {NoStop}%
\bibitem [{\citenamefont {Bertinshaw}\ \emph {et~al.}(2019)\citenamefont
  {Bertinshaw}, \citenamefont {Kim}, \citenamefont {Khaliullin},\ and\
  \citenamefont {Kim}}]{Bertinshaw:19}%
  \BibitemOpen
  \bibfield  {author} {\bibinfo {author} {\bibfnamefont {J.}~\bibnamefont
  {Bertinshaw}}, \bibinfo {author} {\bibfnamefont {Y.~K.}\ \bibnamefont {Kim}},
  \bibinfo {author} {\bibfnamefont {G.}~\bibnamefont {Khaliullin}}, \ and\
  \bibinfo {author} {\bibfnamefont {B.~J.}\ \bibnamefont {Kim}},\ }\href
  {\doibase 10.1146/annurev-conmatphys-031218-013113} {\bibfield  {journal}
  {\bibinfo  {journal} {Annu. Rev. of Condens. Matter Phys.}\ }\textbf
  {\bibinfo {volume} {10}},\ \bibinfo {pages} {315} (\bibinfo {year}
  {2019})}\BibitemShut {NoStop}%
\bibitem [{\citenamefont {Chapon}\ and\ \citenamefont
  {Lovesey}(2011)}]{Chapon:11}%
  \BibitemOpen
  \bibfield  {author} {\bibinfo {author} {\bibfnamefont {L.~C.}\ \bibnamefont
  {Chapon}}\ and\ \bibinfo {author} {\bibfnamefont {S.~W.}\ \bibnamefont
  {Lovesey}},\ }\href {\doibase 10.1088/0953-8984/23/25/252201} {\bibfield
  {journal} {\bibinfo  {journal} {J. Phys.: Condens. Matt.}\ }\textbf {\bibinfo
  {volume} {23}},\ \bibinfo {pages} {252201} (\bibinfo {year}
  {2011})}\BibitemShut {NoStop}%
\bibitem [{\citenamefont {Haskel}\ \emph {et~al.}(2012)\citenamefont {Haskel},
  \citenamefont {Fabbris}, \citenamefont {Zhernenkov}, \citenamefont {Kong},
  \citenamefont {Jin}, \citenamefont {Cao},\ and\ \citenamefont {van
  Veenendaal}}]{Haskel:12}%
  \BibitemOpen
  \bibfield  {author} {\bibinfo {author} {\bibfnamefont {D.}~\bibnamefont
  {Haskel}}, \bibinfo {author} {\bibfnamefont {G.}~\bibnamefont {Fabbris}},
  \bibinfo {author} {\bibfnamefont {M.}~\bibnamefont {Zhernenkov}}, \bibinfo
  {author} {\bibfnamefont {P.~P.}\ \bibnamefont {Kong}}, \bibinfo {author}
  {\bibfnamefont {C.~Q.}\ \bibnamefont {Jin}}, \bibinfo {author} {\bibfnamefont
  {G.}~\bibnamefont {Cao}}, \ and\ \bibinfo {author} {\bibfnamefont
  {M.}~\bibnamefont {van Veenendaal}},\ }\href {\doibase
  10.1103/PhysRevLett.109.027204} {\bibfield  {journal} {\bibinfo  {journal}
  {Phys. Rev. Lett.}\ }\textbf {\bibinfo {volume} {109}},\ \bibinfo {pages}
  {027204} (\bibinfo {year} {2012})}\BibitemShut {NoStop}%
\bibitem [{\citenamefont {Moretti~Sala}\ \emph {et~al.}(2014)\citenamefont
  {Moretti~Sala}, \citenamefont {Boseggia}, \citenamefont {McMorrow},\ and\
  \citenamefont {Monaco}}]{Moretti:14}%
  \BibitemOpen
  \bibfield  {author} {\bibinfo {author} {\bibfnamefont {M.}~\bibnamefont
  {Moretti~Sala}}, \bibinfo {author} {\bibfnamefont {S.}~\bibnamefont
  {Boseggia}}, \bibinfo {author} {\bibfnamefont {D.~F.}\ \bibnamefont
  {McMorrow}}, \ and\ \bibinfo {author} {\bibfnamefont {G.}~\bibnamefont
  {Monaco}},\ }\href {\doibase 10.1103/PhysRevLett.112.026403} {\bibfield
  {journal} {\bibinfo  {journal} {Phys. Rev. Lett.}\ }\textbf {\bibinfo
  {volume} {112}},\ \bibinfo {pages} {026403} (\bibinfo {year}
  {2014})}\BibitemShut {NoStop}%
\bibitem [{\citenamefont {Jeong}\ \emph {et~al.}(2020)\citenamefont {Jeong},
  \citenamefont {Lenz}, \citenamefont {Gukasov}, \citenamefont {Fabr\`eges},
  \citenamefont {Sazonov}, \citenamefont {Hutanu}, \citenamefont {Louat},
  \citenamefont {Bounoua}, \citenamefont {Martins}, \citenamefont {Biermann},
  \citenamefont {Brouet}, \citenamefont {Sidis},\ and\ \citenamefont
  {Bourges}}]{Jeong:20}%
  \BibitemOpen
  \bibfield  {author} {\bibinfo {author} {\bibfnamefont {J.}~\bibnamefont
  {Jeong}}, \bibinfo {author} {\bibfnamefont {B.}~\bibnamefont {Lenz}},
  \bibinfo {author} {\bibfnamefont {A.}~\bibnamefont {Gukasov}}, \bibinfo
  {author} {\bibfnamefont {X.}~\bibnamefont {Fabr\`eges}}, \bibinfo {author}
  {\bibfnamefont {A.}~\bibnamefont {Sazonov}}, \bibinfo {author} {\bibfnamefont
  {V.}~\bibnamefont {Hutanu}}, \bibinfo {author} {\bibfnamefont
  {A.}~\bibnamefont {Louat}}, \bibinfo {author} {\bibfnamefont
  {D.}~\bibnamefont {Bounoua}}, \bibinfo {author} {\bibfnamefont
  {C.}~\bibnamefont {Martins}}, \bibinfo {author} {\bibfnamefont
  {S.}~\bibnamefont {Biermann}}, \bibinfo {author} {\bibfnamefont
  {V.}~\bibnamefont {Brouet}}, \bibinfo {author} {\bibfnamefont
  {Y.}~\bibnamefont {Sidis}}, \ and\ \bibinfo {author} {\bibfnamefont
  {P.}~\bibnamefont {Bourges}},\ }\href {\doibase
  10.1103/PhysRevLett.125.097202} {\bibfield  {journal} {\bibinfo  {journal}
  {Phys. Rev. Lett.}\ }\textbf {\bibinfo {volume} {125}},\ \bibinfo {pages}
  {097202} (\bibinfo {year} {2020})}\BibitemShut {NoStop}%
\bibitem [{\citenamefont {Perkins}\ \emph {et~al.}(2014)\citenamefont
  {Perkins}, \citenamefont {Sizyuk},\ and\ \citenamefont
  {W\"olfle}}]{Perkins:14}%
  \BibitemOpen
  \bibfield  {author} {\bibinfo {author} {\bibfnamefont {N.~B.}\ \bibnamefont
  {Perkins}}, \bibinfo {author} {\bibfnamefont {Y.}~\bibnamefont {Sizyuk}}, \
  and\ \bibinfo {author} {\bibfnamefont {P.}~\bibnamefont {W\"olfle}},\ }\href
  {\doibase 10.1103/PhysRevB.89.035143} {\bibfield  {journal} {\bibinfo
  {journal} {Phys. Rev. B}\ }\textbf {\bibinfo {volume} {89}},\ \bibinfo
  {pages} {035143} (\bibinfo {year} {2014})}\BibitemShut {NoStop}%
\bibitem [{\citenamefont {Crawford}\ \emph {et~al.}(1994)\citenamefont
  {Crawford}, \citenamefont {Subramanian}, \citenamefont {Harlow},
  \citenamefont {Fernandez-Baca}, \citenamefont {Wang},\ and\ \citenamefont
  {Johnston}}]{Crawford:94}%
  \BibitemOpen
  \bibfield  {author} {\bibinfo {author} {\bibfnamefont {M.~K.}\ \bibnamefont
  {Crawford}}, \bibinfo {author} {\bibfnamefont {M.~A.}\ \bibnamefont
  {Subramanian}}, \bibinfo {author} {\bibfnamefont {R.~L.}\ \bibnamefont
  {Harlow}}, \bibinfo {author} {\bibfnamefont {J.~A.}\ \bibnamefont
  {Fernandez-Baca}}, \bibinfo {author} {\bibfnamefont {Z.~R.}\ \bibnamefont
  {Wang}}, \ and\ \bibinfo {author} {\bibfnamefont {D.~C.}\ \bibnamefont
  {Johnston}},\ }\href {\doibase 10.1103/PhysRevB.49.9198} {\bibfield
  {journal} {\bibinfo  {journal} {Phys. Rev. B}\ }\textbf {\bibinfo {volume}
  {49}},\ \bibinfo {pages} {9198} (\bibinfo {year} {1994})}\BibitemShut
  {NoStop}%
\bibitem [{\citenamefont {Huang}\ \emph {et~al.}(1994)\citenamefont {Huang},
  \citenamefont {Soubeyroux}, \citenamefont {Chmaissem}, \citenamefont {Sora},
  \citenamefont {Santoro}, \citenamefont {Cava}, \citenamefont {Krajewski},\
  and\ \citenamefont {Peck}}]{Huang:94}%
  \BibitemOpen
  \bibfield  {author} {\bibinfo {author} {\bibfnamefont {Q.}~\bibnamefont
  {Huang}}, \bibinfo {author} {\bibfnamefont {J.}~\bibnamefont {Soubeyroux}},
  \bibinfo {author} {\bibfnamefont {O.}~\bibnamefont {Chmaissem}}, \bibinfo
  {author} {\bibfnamefont {I.}~\bibnamefont {Sora}}, \bibinfo {author}
  {\bibfnamefont {A.}~\bibnamefont {Santoro}}, \bibinfo {author} {\bibfnamefont
  {R.}~\bibnamefont {Cava}}, \bibinfo {author} {\bibfnamefont {J.}~\bibnamefont
  {Krajewski}}, \ and\ \bibinfo {author} {\bibfnamefont {W.}~\bibnamefont
  {Peck}},\ }\href {\doibase https://doi.org/10.1006/jssc.1994.1316} {\bibfield
   {journal} {\bibinfo  {journal} {J. Solid State Chem.}\ }\textbf {\bibinfo
  {volume} {112}},\ \bibinfo {pages} {355} (\bibinfo {year}
  {1994})}\BibitemShut {NoStop}%
\bibitem [{\citenamefont {Ge}\ \emph {et~al.}(2011)\citenamefont {Ge},
  \citenamefont {Qi}, \citenamefont {Korneta}, \citenamefont {De~Long},
  \citenamefont {Schlottmann}, \citenamefont {Crummett},\ and\ \citenamefont
  {Cao}}]{Ge:11}%
  \BibitemOpen
  \bibfield  {author} {\bibinfo {author} {\bibfnamefont {M.}~\bibnamefont
  {Ge}}, \bibinfo {author} {\bibfnamefont {T.~F.}\ \bibnamefont {Qi}}, \bibinfo
  {author} {\bibfnamefont {O.~B.}\ \bibnamefont {Korneta}}, \bibinfo {author}
  {\bibfnamefont {D.~E.}\ \bibnamefont {De~Long}}, \bibinfo {author}
  {\bibfnamefont {P.}~\bibnamefont {Schlottmann}}, \bibinfo {author}
  {\bibfnamefont {W.~P.}\ \bibnamefont {Crummett}}, \ and\ \bibinfo {author}
  {\bibfnamefont {G.}~\bibnamefont {Cao}},\ }\href {\doibase
  10.1103/PhysRevB.84.100402} {\bibfield  {journal} {\bibinfo  {journal} {Phys.
  Rev. B}\ }\textbf {\bibinfo {volume} {84}},\ \bibinfo {pages} {100402}
  (\bibinfo {year} {2011})}\BibitemShut {NoStop}%
\bibitem [{\citenamefont {Ye}\ \emph {et~al.}(2013)\citenamefont {Ye},
  \citenamefont {Chi}, \citenamefont {Chakoumakos}, \citenamefont
  {Fernandez-Baca}, \citenamefont {Qi},\ and\ \citenamefont {Cao}}]{Ye:13}%
  \BibitemOpen
  \bibfield  {author} {\bibinfo {author} {\bibfnamefont {F.}~\bibnamefont
  {Ye}}, \bibinfo {author} {\bibfnamefont {S.}~\bibnamefont {Chi}}, \bibinfo
  {author} {\bibfnamefont {B.~C.}\ \bibnamefont {Chakoumakos}}, \bibinfo
  {author} {\bibfnamefont {J.~A.}\ \bibnamefont {Fernandez-Baca}}, \bibinfo
  {author} {\bibfnamefont {T.}~\bibnamefont {Qi}}, \ and\ \bibinfo {author}
  {\bibfnamefont {G.}~\bibnamefont {Cao}},\ }\href {\doibase
  10.1103/PhysRevB.87.140406} {\bibfield  {journal} {\bibinfo  {journal} {Phys.
  Rev. B}\ }\textbf {\bibinfo {volume} {87}},\ \bibinfo {pages} {140406}
  (\bibinfo {year} {2013})}\BibitemShut {NoStop}%
\bibitem [{\citenamefont {Dhital}\ \emph {et~al.}(2013)\citenamefont {Dhital},
  \citenamefont {Hogan}, \citenamefont {Yamani}, \citenamefont {de~la Cruz},
  \citenamefont {Chen}, \citenamefont {Khadka}, \citenamefont {Ren},\ and\
  \citenamefont {Wilson}}]{Dhital:13}%
  \BibitemOpen
  \bibfield  {author} {\bibinfo {author} {\bibfnamefont {C.}~\bibnamefont
  {Dhital}}, \bibinfo {author} {\bibfnamefont {T.}~\bibnamefont {Hogan}},
  \bibinfo {author} {\bibfnamefont {Z.}~\bibnamefont {Yamani}}, \bibinfo
  {author} {\bibfnamefont {C.}~\bibnamefont {de~la Cruz}}, \bibinfo {author}
  {\bibfnamefont {X.}~\bibnamefont {Chen}}, \bibinfo {author} {\bibfnamefont
  {S.}~\bibnamefont {Khadka}}, \bibinfo {author} {\bibfnamefont
  {Z.}~\bibnamefont {Ren}}, \ and\ \bibinfo {author} {\bibfnamefont {S.~D.}\
  \bibnamefont {Wilson}},\ }\href {\doibase 10.1103/PhysRevB.87.144405}
  {\bibfield  {journal} {\bibinfo  {journal} {Phys. Rev. B}\ }\textbf {\bibinfo
  {volume} {87}},\ \bibinfo {pages} {144405} (\bibinfo {year}
  {2013})}\BibitemShut {NoStop}%
\bibitem [{\citenamefont {Sung}\ \emph {et~al.}(2016)\citenamefont {Sung},
  \citenamefont {Gretarsson}, \citenamefont {Proepper}, \citenamefont {Porras},
  \citenamefont {Tacon}, \citenamefont {Boris}, \citenamefont {Keimer},\ and\
  \citenamefont {Kim}}]{Sung:16}%
  \BibitemOpen
  \bibfield  {author} {\bibinfo {author} {\bibfnamefont {N.~H.}\ \bibnamefont
  {Sung}}, \bibinfo {author} {\bibfnamefont {H.}~\bibnamefont {Gretarsson}},
  \bibinfo {author} {\bibfnamefont {D.}~\bibnamefont {Proepper}}, \bibinfo
  {author} {\bibfnamefont {J.}~\bibnamefont {Porras}}, \bibinfo {author}
  {\bibfnamefont {M.~L.}\ \bibnamefont {Tacon}}, \bibinfo {author}
  {\bibfnamefont {A.~V.}\ \bibnamefont {Boris}}, \bibinfo {author}
  {\bibfnamefont {B.}~\bibnamefont {Keimer}}, \ and\ \bibinfo {author}
  {\bibfnamefont {B.~J.}\ \bibnamefont {Kim}},\ }\href {\doibase
  10.1080/14786435.2015.1134835} {\bibfield  {journal} {\bibinfo  {journal}
  {Phil. Mag.}\ }\textbf {\bibinfo {volume} {96}},\ \bibinfo {pages} {413}
  (\bibinfo {year} {2016})}\BibitemShut {NoStop}%
\bibitem [{\citenamefont {Chikara}\ \emph {et~al.}(2009)\citenamefont
  {Chikara}, \citenamefont {Korneta}, \citenamefont {Crummett}, \citenamefont
  {DeLong}, \citenamefont {Schlottmann},\ and\ \citenamefont
  {Cao}}]{Chikara:09}%
  \BibitemOpen
  \bibfield  {author} {\bibinfo {author} {\bibfnamefont {S.}~\bibnamefont
  {Chikara}}, \bibinfo {author} {\bibfnamefont {O.}~\bibnamefont {Korneta}},
  \bibinfo {author} {\bibfnamefont {W.~P.}\ \bibnamefont {Crummett}}, \bibinfo
  {author} {\bibfnamefont {L.~E.}\ \bibnamefont {DeLong}}, \bibinfo {author}
  {\bibfnamefont {P.}~\bibnamefont {Schlottmann}}, \ and\ \bibinfo {author}
  {\bibfnamefont {G.}~\bibnamefont {Cao}},\ }\href {\doibase
  10.1103/PhysRevB.80.140407} {\bibfield  {journal} {\bibinfo  {journal} {Phys.
  Rev. B}\ }\textbf {\bibinfo {volume} {80}},\ \bibinfo {pages} {140407}
  (\bibinfo {year} {2009})}\BibitemShut {NoStop}%
\bibitem [{\citenamefont {Li}\ \emph {et~al.}(2013)\citenamefont {Li},
  \citenamefont {Kong}, \citenamefont {Qi}, \citenamefont {Jin}, \citenamefont
  {Yuan}, \citenamefont {DeLong}, \citenamefont {Schlottmann},\ and\
  \citenamefont {Cao}}]{Li:13}%
  \BibitemOpen
  \bibfield  {author} {\bibinfo {author} {\bibfnamefont {L.}~\bibnamefont
  {Li}}, \bibinfo {author} {\bibfnamefont {P.~P.}\ \bibnamefont {Kong}},
  \bibinfo {author} {\bibfnamefont {T.~F.}\ \bibnamefont {Qi}}, \bibinfo
  {author} {\bibfnamefont {C.~Q.}\ \bibnamefont {Jin}}, \bibinfo {author}
  {\bibfnamefont {S.~J.}\ \bibnamefont {Yuan}}, \bibinfo {author}
  {\bibfnamefont {L.~E.}\ \bibnamefont {DeLong}}, \bibinfo {author}
  {\bibfnamefont {P.}~\bibnamefont {Schlottmann}}, \ and\ \bibinfo {author}
  {\bibfnamefont {G.}~\bibnamefont {Cao}},\ }\href {\doibase
  10.1103/PhysRevB.87.235127} {\bibfield  {journal} {\bibinfo  {journal} {Phys.
  Rev. B}\ }\textbf {\bibinfo {volume} {87}},\ \bibinfo {pages} {235127}
  (\bibinfo {year} {2013})}\BibitemShut {NoStop}%
\bibitem [{\citenamefont {Katsura}\ \emph {et~al.}(2005)\citenamefont
  {Katsura}, \citenamefont {Nagaosa},\ and\ \citenamefont
  {Balatsky}}]{Katsura:05}%
  \BibitemOpen
  \bibfield  {author} {\bibinfo {author} {\bibfnamefont {H.}~\bibnamefont
  {Katsura}}, \bibinfo {author} {\bibfnamefont {N.}~\bibnamefont {Nagaosa}}, \
  and\ \bibinfo {author} {\bibfnamefont {A.~V.}\ \bibnamefont {Balatsky}},\
  }\href {\doibase 10.1103/PhysRevLett.95.057205} {\bibfield  {journal}
  {\bibinfo  {journal} {Phys. Rev. Lett.}\ }\textbf {\bibinfo {volume} {95}},\
  \bibinfo {pages} {057205} (\bibinfo {year} {2005})}\BibitemShut {NoStop}%
\bibitem [{\citenamefont {Nagaosa}(2008)}]{Nagaosa:08}%
  \BibitemOpen
  \bibfield  {author} {\bibinfo {author} {\bibfnamefont {N.}~\bibnamefont
  {Nagaosa}},\ }\href@noop {} {\bibfield  {journal} {\bibinfo  {journal} {J.
  Phys.: Condens. Matt.}\ }\textbf {\bibinfo {volume} {20}},\ \bibinfo {pages}
  {434207} (\bibinfo {year} {2008})}\BibitemShut {NoStop}%
\bibitem [{\citenamefont {Sergienko}\ and\ \citenamefont
  {Dagotto}(2006)}]{Sergienko:06}%
  \BibitemOpen
  \bibfield  {author} {\bibinfo {author} {\bibfnamefont {I.~A.}\ \bibnamefont
  {Sergienko}}\ and\ \bibinfo {author} {\bibfnamefont {E.}~\bibnamefont
  {Dagotto}},\ }\href {\doibase 10.1103/PhysRevB.73.094434} {\bibfield
  {journal} {\bibinfo  {journal} {Phys. Rev. B}\ }\textbf {\bibinfo {volume}
  {73}},\ \bibinfo {pages} {094434} (\bibinfo {year} {2006})}\BibitemShut
  {NoStop}%
\bibitem [{\citenamefont {Momma}\ and\ \citenamefont {Izumi}(2011)}]{Vesta}%
  \BibitemOpen
  \bibfield  {author} {\bibinfo {author} {\bibfnamefont {K.}~\bibnamefont
  {Momma}}\ and\ \bibinfo {author} {\bibfnamefont {F.}~\bibnamefont {Izumi}},\
  }\href {\doibase 10.1107/S0021889811038970} {\bibfield  {journal} {\bibinfo
  {journal} {J. Appl. Crystallogr.}\ }\textbf {\bibinfo {volume} {44}},\
  \bibinfo {pages} {1272} (\bibinfo {year} {2011})}\BibitemShut {NoStop}%
\bibitem [{\citenamefont {Miyazaki}\ \emph {et~al.}(2015)\citenamefont
  {Miyazaki}, \citenamefont {Kadono}, \citenamefont {Hiraishi}, \citenamefont
  {Koda}, \citenamefont {Kojima}, \citenamefont {Ohashi}, \citenamefont
  {Takayama},\ and\ \citenamefont {Takagi}}]{Miyazaki:15}%
  \BibitemOpen
  \bibfield  {author} {\bibinfo {author} {\bibfnamefont {M.}~\bibnamefont
  {Miyazaki}}, \bibinfo {author} {\bibfnamefont {R.}~\bibnamefont {Kadono}},
  \bibinfo {author} {\bibfnamefont {M.}~\bibnamefont {Hiraishi}}, \bibinfo
  {author} {\bibfnamefont {A.}~\bibnamefont {Koda}}, \bibinfo {author}
  {\bibfnamefont {K.~M.}\ \bibnamefont {Kojima}}, \bibinfo {author}
  {\bibfnamefont {K.}~\bibnamefont {Ohashi}}, \bibinfo {author} {\bibfnamefont
  {T.}~\bibnamefont {Takayama}}, \ and\ \bibinfo {author} {\bibfnamefont
  {H.}~\bibnamefont {Takagi}},\ }\href {\doibase 10.1103/PhysRevB.91.155113}
  {\bibfield  {journal} {\bibinfo  {journal} {Phys. Rev. B}\ }\textbf {\bibinfo
  {volume} {91}},\ \bibinfo {pages} {155113} (\bibinfo {year}
  {2015})}\BibitemShut {NoStop}%
\bibitem [{\citenamefont {Ishii}\ \emph {et~al.}(2020)\citenamefont {Ishii},
  \citenamefont {Horio}, \citenamefont {Noda}, \citenamefont {Hiraishi},
  \citenamefont {Okabe}, \citenamefont {Miyazaki}, \citenamefont {Takeshita},
  \citenamefont {Koda}, \citenamefont {Kojima}, \citenamefont {Kadono},
  \citenamefont {Sagayama}, \citenamefont {Nakao}, \citenamefont {Murakami},\
  and\ \citenamefont {Kimura}}]{Ishii:20}%
  \BibitemOpen
  \bibfield  {author} {\bibinfo {author} {\bibfnamefont {Y.}~\bibnamefont
  {Ishii}}, \bibinfo {author} {\bibfnamefont {S.}~\bibnamefont {Horio}},
  \bibinfo {author} {\bibfnamefont {Y.}~\bibnamefont {Noda}}, \bibinfo {author}
  {\bibfnamefont {M.}~\bibnamefont {Hiraishi}}, \bibinfo {author}
  {\bibfnamefont {H.}~\bibnamefont {Okabe}}, \bibinfo {author} {\bibfnamefont
  {M.}~\bibnamefont {Miyazaki}}, \bibinfo {author} {\bibfnamefont
  {S.}~\bibnamefont {Takeshita}}, \bibinfo {author} {\bibfnamefont
  {A.}~\bibnamefont {Koda}}, \bibinfo {author} {\bibfnamefont {K.~M.}\
  \bibnamefont {Kojima}}, \bibinfo {author} {\bibfnamefont {R.}~\bibnamefont
  {Kadono}}, \bibinfo {author} {\bibfnamefont {H.}~\bibnamefont {Sagayama}},
  \bibinfo {author} {\bibfnamefont {H.}~\bibnamefont {Nakao}}, \bibinfo
  {author} {\bibfnamefont {Y.}~\bibnamefont {Murakami}}, \ and\ \bibinfo
  {author} {\bibfnamefont {H.}~\bibnamefont {Kimura}},\ }\href {\doibase
  10.1103/PhysRevB.101.224436} {\bibfield  {journal} {\bibinfo  {journal}
  {Phys. Rev. B}\ }\textbf {\bibinfo {volume} {101}},\ \bibinfo {pages}
  {224436} (\bibinfo {year} {2020})}\BibitemShut {NoStop}%
\bibitem [{\citenamefont {Dehn}\ \emph {et~al.}(2020)\citenamefont {Dehn},
  \citenamefont {Shenton}, \citenamefont {Holenstein}, \citenamefont {Meier},
  \citenamefont {Arseneau}, \citenamefont {Cortie}, \citenamefont {Hitti},
  \citenamefont {Fang}, \citenamefont {MacFarlane}, \citenamefont {McFadden},
  \citenamefont {Morris}, \citenamefont {Salman}, \citenamefont {Luetkens},
  \citenamefont {Spaldin}, \citenamefont {Fechner},\ and\ \citenamefont
  {Kiefl}}]{Dehn:20}%
  \BibitemOpen
  \bibfield  {author} {\bibinfo {author} {\bibfnamefont {M.~H.}\ \bibnamefont
  {Dehn}}, \bibinfo {author} {\bibfnamefont {J.~K.}\ \bibnamefont {Shenton}},
  \bibinfo {author} {\bibfnamefont {S.}~\bibnamefont {Holenstein}}, \bibinfo
  {author} {\bibfnamefont {Q.~N.}\ \bibnamefont {Meier}}, \bibinfo {author}
  {\bibfnamefont {D.~J.}\ \bibnamefont {Arseneau}}, \bibinfo {author}
  {\bibfnamefont {D.~L.}\ \bibnamefont {Cortie}}, \bibinfo {author}
  {\bibfnamefont {B.}~\bibnamefont {Hitti}}, \bibinfo {author} {\bibfnamefont
  {A.~C.~Y.}\ \bibnamefont {Fang}}, \bibinfo {author} {\bibfnamefont {W.~A.}\
  \bibnamefont {MacFarlane}}, \bibinfo {author} {\bibfnamefont {R.~M.~L.}\
  \bibnamefont {McFadden}}, \bibinfo {author} {\bibfnamefont {G.~D.}\
  \bibnamefont {Morris}}, \bibinfo {author} {\bibfnamefont {Z.}~\bibnamefont
  {Salman}}, \bibinfo {author} {\bibfnamefont {H.}~\bibnamefont {Luetkens}},
  \bibinfo {author} {\bibfnamefont {N.~A.}\ \bibnamefont {Spaldin}}, \bibinfo
  {author} {\bibfnamefont {M.}~\bibnamefont {Fechner}}, \ and\ \bibinfo
  {author} {\bibfnamefont {R.~F.}\ \bibnamefont {Kiefl}},\ }\href {\doibase
  10.1103/PhysRevX.10.011036} {\bibfield  {journal} {\bibinfo  {journal} {Phys.
  Rev. X}\ }\textbf {\bibinfo {volume} {10}},\ \bibinfo {pages} {011036}
  (\bibinfo {year} {2020})}\BibitemShut {NoStop}%
\bibitem [{\citenamefont {Dehn}\ \emph {et~al.}(2021)\citenamefont {Dehn},
  \citenamefont {Shenton}, \citenamefont {Arseneau}, \citenamefont
  {MacFarlane}, \citenamefont {Morris}, \citenamefont {Maign\'e}, \citenamefont
  {Spaldin},\ and\ \citenamefont {Kiefl}}]{Dehn:21}%
  \BibitemOpen
  \bibfield  {author} {\bibinfo {author} {\bibfnamefont {M.~H.}\ \bibnamefont
  {Dehn}}, \bibinfo {author} {\bibfnamefont {J.~K.}\ \bibnamefont {Shenton}},
  \bibinfo {author} {\bibfnamefont {D.~J.}\ \bibnamefont {Arseneau}}, \bibinfo
  {author} {\bibfnamefont {W.~A.}\ \bibnamefont {MacFarlane}}, \bibinfo
  {author} {\bibfnamefont {G.~D.}\ \bibnamefont {Morris}}, \bibinfo {author}
  {\bibfnamefont {A.}~\bibnamefont {Maign\'e}}, \bibinfo {author}
  {\bibfnamefont {N.~A.}\ \bibnamefont {Spaldin}}, \ and\ \bibinfo {author}
  {\bibfnamefont {R.~F.}\ \bibnamefont {Kiefl}},\ }\href {\doibase
  10.1103/PhysRevLett.126.037202} {\bibfield  {journal} {\bibinfo  {journal}
  {Phys. Rev. Lett.}\ }\textbf {\bibinfo {volume} {126}},\ \bibinfo {pages}
  {037202} (\bibinfo {year} {2021})}\BibitemShut {NoStop}%
\bibitem [{\citenamefont {Koide}\ \emph {et~al.}(2001)\citenamefont {Koide},
  \citenamefont {Miyauchi}, \citenamefont {Okamoto}, \citenamefont {Shidara},
  \citenamefont {Sekine}, \citenamefont {Saitoh}, \citenamefont {Fujimori},
  \citenamefont {Fukutani}, \citenamefont {Takano},\ and\ \citenamefont
  {Takeda}}]{Koide:01}%
  \BibitemOpen
  \bibfield  {author} {\bibinfo {author} {\bibfnamefont {T.}~\bibnamefont
  {Koide}}, \bibinfo {author} {\bibfnamefont {H.}~\bibnamefont {Miyauchi}},
  \bibinfo {author} {\bibfnamefont {J.}~\bibnamefont {Okamoto}}, \bibinfo
  {author} {\bibfnamefont {T.}~\bibnamefont {Shidara}}, \bibinfo {author}
  {\bibfnamefont {T.}~\bibnamefont {Sekine}}, \bibinfo {author} {\bibfnamefont
  {T.}~\bibnamefont {Saitoh}}, \bibinfo {author} {\bibfnamefont
  {A.}~\bibnamefont {Fujimori}}, \bibinfo {author} {\bibfnamefont
  {H.}~\bibnamefont {Fukutani}}, \bibinfo {author} {\bibfnamefont
  {M.}~\bibnamefont {Takano}}, \ and\ \bibinfo {author} {\bibfnamefont
  {Y.}~\bibnamefont {Takeda}},\ }\href {\doibase 10.1103/PhysRevLett.87.246404}
  {\bibfield  {journal} {\bibinfo  {journal} {Phys. Rev. Lett.}\ }\textbf
  {\bibinfo {volume} {87}},\ \bibinfo {pages} {246404} (\bibinfo {year}
  {2001})}\BibitemShut {NoStop}%
\bibitem [{\citenamefont {Okamoto}\ \emph {et~al.}(2000)\citenamefont
  {Okamoto}, \citenamefont {Miyauchi}, \citenamefont {Sekine}, \citenamefont
  {Shidara}, \citenamefont {Koide}, \citenamefont {Amemiya}, \citenamefont
  {Fujimori}, \citenamefont {Saitoh}, \citenamefont {Tanaka}, \citenamefont
  {Takeda},\ and\ \citenamefont {Takano}}]{Okamoto:00}%
  \BibitemOpen
  \bibfield  {author} {\bibinfo {author} {\bibfnamefont {J.}~\bibnamefont
  {Okamoto}}, \bibinfo {author} {\bibfnamefont {H.}~\bibnamefont {Miyauchi}},
  \bibinfo {author} {\bibfnamefont {T.}~\bibnamefont {Sekine}}, \bibinfo
  {author} {\bibfnamefont {T.}~\bibnamefont {Shidara}}, \bibinfo {author}
  {\bibfnamefont {T.}~\bibnamefont {Koide}}, \bibinfo {author} {\bibfnamefont
  {K.}~\bibnamefont {Amemiya}}, \bibinfo {author} {\bibfnamefont
  {A.}~\bibnamefont {Fujimori}}, \bibinfo {author} {\bibfnamefont
  {T.}~\bibnamefont {Saitoh}}, \bibinfo {author} {\bibfnamefont
  {A.}~\bibnamefont {Tanaka}}, \bibinfo {author} {\bibfnamefont
  {Y.}~\bibnamefont {Takeda}}, \ and\ \bibinfo {author} {\bibfnamefont
  {M.}~\bibnamefont {Takano}},\ }\href {\doibase 10.1103/PhysRevB.62.4455}
  {\bibfield  {journal} {\bibinfo  {journal} {Phys. Rev. B}\ }\textbf {\bibinfo
  {volume} {62}},\ \bibinfo {pages} {4455} (\bibinfo {year}
  {2000})}\BibitemShut {NoStop}%
\bibitem [{\citenamefont {Medling}\ \emph {et~al.}(2012)\citenamefont
  {Medling}, \citenamefont {Lee}, \citenamefont {Zheng}, \citenamefont
  {Mitchell}, \citenamefont {Freeland}, \citenamefont {Harmon},\ and\
  \citenamefont {Bridges}}]{Medling:12}%
  \BibitemOpen
  \bibfield  {author} {\bibinfo {author} {\bibfnamefont {S.}~\bibnamefont
  {Medling}}, \bibinfo {author} {\bibfnamefont {Y.}~\bibnamefont {Lee}},
  \bibinfo {author} {\bibfnamefont {H.}~\bibnamefont {Zheng}}, \bibinfo
  {author} {\bibfnamefont {J.~F.}\ \bibnamefont {Mitchell}}, \bibinfo {author}
  {\bibfnamefont {J.~W.}\ \bibnamefont {Freeland}}, \bibinfo {author}
  {\bibfnamefont {B.~N.}\ \bibnamefont {Harmon}}, \ and\ \bibinfo {author}
  {\bibfnamefont {F.}~\bibnamefont {Bridges}},\ }\href {\doibase
  10.1103/PhysRevLett.109.157204} {\bibfield  {journal} {\bibinfo  {journal}
  {Phys. Rev. Lett.}\ }\textbf {\bibinfo {volume} {109}},\ \bibinfo {pages}
  {157204} (\bibinfo {year} {2012})}\BibitemShut {NoStop}%
\bibitem [{\citenamefont {Goering}\ \emph {et~al.}(2002)\citenamefont
  {Goering}, \citenamefont {Bayer}, \citenamefont {Gold}, \citenamefont
  {Sch\"utz}, \citenamefont {Rabe}, \citenamefont {R\"udiger},\ and\
  \citenamefont {G\"untherodt}}]{Goering:02}%
  \BibitemOpen
  \bibfield  {author} {\bibinfo {author} {\bibfnamefont {E.}~\bibnamefont
  {Goering}}, \bibinfo {author} {\bibfnamefont {A.}~\bibnamefont {Bayer}},
  \bibinfo {author} {\bibfnamefont {S.}~\bibnamefont {Gold}}, \bibinfo {author}
  {\bibfnamefont {G.}~\bibnamefont {Sch\"utz}}, \bibinfo {author}
  {\bibfnamefont {M.}~\bibnamefont {Rabe}}, \bibinfo {author} {\bibfnamefont
  {U.}~\bibnamefont {R\"udiger}}, \ and\ \bibinfo {author} {\bibfnamefont
  {G.}~\bibnamefont {G\"untherodt}},\ }\href {\doibase
  10.1209/epl/i2002-00103-6} {\bibfield  {journal} {\bibinfo  {journal}
  {Europhys. Lett.}\ }\textbf {\bibinfo {volume} {58}},\ \bibinfo {pages} {906}
  (\bibinfo {year} {2002})}\BibitemShut {NoStop}%
\bibitem [{\citenamefont {Amemiya}\ \emph {et~al.}(2010)\citenamefont
  {Amemiya}, \citenamefont {Toyoshima}, \citenamefont {Kikuchi}, \citenamefont
  {Kosuge}, \citenamefont {Nigorikawa}, \citenamefont {Sumii},\ and\
  \citenamefont {Ito}}]{Amemiya:10}%
  \BibitemOpen
  \bibfield  {author} {\bibinfo {author} {\bibfnamefont {K.}~\bibnamefont
  {Amemiya}}, \bibinfo {author} {\bibfnamefont {A.}~\bibnamefont {Toyoshima}},
  \bibinfo {author} {\bibfnamefont {T.}~\bibnamefont {Kikuchi}}, \bibinfo
  {author} {\bibfnamefont {T.}~\bibnamefont {Kosuge}}, \bibinfo {author}
  {\bibfnamefont {K.}~\bibnamefont {Nigorikawa}}, \bibinfo {author}
  {\bibfnamefont {R.}~\bibnamefont {Sumii}}, \ and\ \bibinfo {author}
  {\bibfnamefont {K.}~\bibnamefont {Ito}},\ }\href {\doibase 10.1063/1.3463193}
  {\bibfield  {journal} {\bibinfo  {journal} {AIP Conf. Proc.}\ }\textbf
  {\bibinfo {volume} {1234}},\ \bibinfo {pages} {295} (\bibinfo {year}
  {2010})}\BibitemShut {NoStop}%
\bibitem [{\citenamefont {Amemiya}\ \emph {et~al.}(2013)\citenamefont
  {Amemiya}, \citenamefont {Sakamaki}, \citenamefont {Koide}, \citenamefont
  {Ito}, \citenamefont {.Tsuchiya}, \citenamefont {Harada}, \citenamefont
  {Aoto}, \citenamefont {Shioya}, \citenamefont {Obina}, \citenamefont
  {Yamamoto},\ and\ \citenamefont {Kobayashi}}]{Amemiya:13}%
  \BibitemOpen
  \bibfield  {author} {\bibinfo {author} {\bibfnamefont {K.}~\bibnamefont
  {Amemiya}}, \bibinfo {author} {\bibfnamefont {M.}~\bibnamefont {Sakamaki}},
  \bibinfo {author} {\bibfnamefont {T.}~\bibnamefont {Koide}}, \bibinfo
  {author} {\bibfnamefont {K.}~\bibnamefont {Ito}}, \bibinfo {author}
  {\bibfnamefont {K.}~\bibnamefont {.Tsuchiya}}, \bibinfo {author}
  {\bibfnamefont {K.}~\bibnamefont {Harada}}, \bibinfo {author} {\bibfnamefont
  {T.}~\bibnamefont {Aoto}}, \bibinfo {author} {\bibfnamefont {T.}~\bibnamefont
  {Shioya}}, \bibinfo {author} {\bibfnamefont {T.}~\bibnamefont {Obina}},
  \bibinfo {author} {\bibfnamefont {S.}~\bibnamefont {Yamamoto}}, \ and\
  \bibinfo {author} {\bibfnamefont {Y.}~\bibnamefont {Kobayashi}},\ }\href@noop
  {} {\bibfield  {journal} {\bibinfo  {journal} {J. Phys.: Conf. Ser.}\
  }\textbf {\bibinfo {volume} {425}},\ \bibinfo {pages} {152015} (\bibinfo
  {year} {2013})}\BibitemShut {NoStop}%
\bibitem [{\citenamefont {Moon}\ \emph {et~al.}(2006)\citenamefont {Moon},
  \citenamefont {Kim}, \citenamefont {Kim}, \citenamefont {Lee}, \citenamefont
  {Kim}, \citenamefont {Park}, \citenamefont {Kim}, \citenamefont {Oh},
  \citenamefont {Nakatsuji}, \citenamefont {Maeno}, \citenamefont {Nagai},
  \citenamefont {Ikeda}, \citenamefont {Cao},\ and\ \citenamefont
  {Noh}}]{Moon:06}%
  \BibitemOpen
  \bibfield  {author} {\bibinfo {author} {\bibfnamefont {S.~J.}\ \bibnamefont
  {Moon}}, \bibinfo {author} {\bibfnamefont {M.~W.}\ \bibnamefont {Kim}},
  \bibinfo {author} {\bibfnamefont {K.~W.}\ \bibnamefont {Kim}}, \bibinfo
  {author} {\bibfnamefont {Y.~S.}\ \bibnamefont {Lee}}, \bibinfo {author}
  {\bibfnamefont {J.-Y.}\ \bibnamefont {Kim}}, \bibinfo {author} {\bibfnamefont
  {J.-H.}\ \bibnamefont {Park}}, \bibinfo {author} {\bibfnamefont {B.~J.}\
  \bibnamefont {Kim}}, \bibinfo {author} {\bibfnamefont {S.-J.}\ \bibnamefont
  {Oh}}, \bibinfo {author} {\bibfnamefont {S.}~\bibnamefont {Nakatsuji}},
  \bibinfo {author} {\bibfnamefont {Y.}~\bibnamefont {Maeno}}, \bibinfo
  {author} {\bibfnamefont {I.}~\bibnamefont {Nagai}}, \bibinfo {author}
  {\bibfnamefont {S.~I.}\ \bibnamefont {Ikeda}}, \bibinfo {author}
  {\bibfnamefont {G.}~\bibnamefont {Cao}}, \ and\ \bibinfo {author}
  {\bibfnamefont {T.~W.}\ \bibnamefont {Noh}},\ }\href {\doibase
  10.1103/PhysRevB.74.113104} {\bibfield  {journal} {\bibinfo  {journal} {Phys.
  Rev. B}\ }\textbf {\bibinfo {volume} {74}},\ \bibinfo {pages} {113104}
  (\bibinfo {year} {2006})}\BibitemShut {NoStop}%
\bibitem [{\citenamefont {Liu}\ \emph {et~al.}(2015)\citenamefont {Liu},
  \citenamefont {Dean}, \citenamefont {Liu}, \citenamefont {Chiuzb{\u a}ian},
  \citenamefont {Jaouen}, \citenamefont {Nicolaou}, \citenamefont {Yin},
  \citenamefont {Serrao}, \citenamefont {Ramesh}, \citenamefont {Ding},\ and\
  \citenamefont {Hill}}]{Liu:15}%
  \BibitemOpen
  \bibfield  {author} {\bibinfo {author} {\bibfnamefont {X.}~\bibnamefont
  {Liu}}, \bibinfo {author} {\bibfnamefont {M.~P.~M.}\ \bibnamefont {Dean}},
  \bibinfo {author} {\bibfnamefont {J.}~\bibnamefont {Liu}}, \bibinfo {author}
  {\bibfnamefont {S.~G.}\ \bibnamefont {Chiuzb{\u a}ian}}, \bibinfo {author}
  {\bibfnamefont {N.}~\bibnamefont {Jaouen}}, \bibinfo {author} {\bibfnamefont
  {A.}~\bibnamefont {Nicolaou}}, \bibinfo {author} {\bibfnamefont {W.~G.}\
  \bibnamefont {Yin}}, \bibinfo {author} {\bibfnamefont {C.~R.}\ \bibnamefont
  {Serrao}}, \bibinfo {author} {\bibfnamefont {R.}~\bibnamefont {Ramesh}},
  \bibinfo {author} {\bibfnamefont {H.}~\bibnamefont {Ding}}, \ and\ \bibinfo
  {author} {\bibfnamefont {J.~P.}\ \bibnamefont {Hill}},\ }\href {\doibase
  10.1088/0953-8984/27/20/202202} {\bibfield  {journal} {\bibinfo  {journal}
  {J. Phys.: Condens. Matt.}\ }\textbf {\bibinfo {volume} {27}},\ \bibinfo
  {pages} {202202} (\bibinfo {year} {2015})}\BibitemShut {NoStop}%
\bibitem [{\citenamefont {Ilakovac}\ \emph {et~al.}(2019)\citenamefont
  {Ilakovac}, \citenamefont {Louat}, \citenamefont {Nicolaou}, \citenamefont
  {Rueff}, \citenamefont {Joly},\ and\ \citenamefont {Brouet}}]{Ilakovac:19}%
  \BibitemOpen
  \bibfield  {author} {\bibinfo {author} {\bibfnamefont {V.}~\bibnamefont
  {Ilakovac}}, \bibinfo {author} {\bibfnamefont {A.}~\bibnamefont {Louat}},
  \bibinfo {author} {\bibfnamefont {A.}~\bibnamefont {Nicolaou}}, \bibinfo
  {author} {\bibfnamefont {J.-P.}\ \bibnamefont {Rueff}}, \bibinfo {author}
  {\bibfnamefont {Y.}~\bibnamefont {Joly}}, \ and\ \bibinfo {author}
  {\bibfnamefont {V.}~\bibnamefont {Brouet}},\ }\href {\doibase
  10.1103/PhysRevB.99.035149} {\bibfield  {journal} {\bibinfo  {journal} {Phys.
  Rev. B}\ }\textbf {\bibinfo {volume} {99}},\ \bibinfo {pages} {035149}
  (\bibinfo {year} {2019})}\BibitemShut {NoStop}%
\bibitem [{\citenamefont {Bhandari}\ \emph {et~al.}(2019)\citenamefont
  {Bhandari}, \citenamefont {Popovi{\'c}},\ and\ \citenamefont
  {Satpathy}}]{Bhandari:19}%
  \BibitemOpen
  \bibfield  {author} {\bibinfo {author} {\bibfnamefont {C.}~\bibnamefont
  {Bhandari}}, \bibinfo {author} {\bibfnamefont {Z.~S.}\ \bibnamefont
  {Popovi{\'c}}}, \ and\ \bibinfo {author} {\bibfnamefont {S.}~\bibnamefont
  {Satpathy}},\ }\href {\doibase 10.1088/1367-2630/aaff1b} {\bibfield
  {journal} {\bibinfo  {journal} {New J. Phys.}\ }\textbf {\bibinfo {volume}
  {21}},\ \bibinfo {pages} {013036} (\bibinfo {year} {2019})}\BibitemShut
  {NoStop}%
\bibitem [{\citenamefont {Lu}\ \emph {et~al.}(2018)\citenamefont {Lu},
  \citenamefont {Olalde-Velasco}, \citenamefont {Huang}, \citenamefont
  {Bisogni}, \citenamefont {Pelliciari}, \citenamefont {Fatale}, \citenamefont
  {Dantz}, \citenamefont {Vale}, \citenamefont {Hunter}, \citenamefont {Chang},
  \citenamefont {Strocov}, \citenamefont {Perry}, \citenamefont {Grioni},
  \citenamefont {McMorrow}, \citenamefont {R\o{}nnow},\ and\ \citenamefont
  {Schmitt}}]{Lu:18}%
  \BibitemOpen
  \bibfield  {author} {\bibinfo {author} {\bibfnamefont {X.}~\bibnamefont
  {Lu}}, \bibinfo {author} {\bibfnamefont {P.}~\bibnamefont {Olalde-Velasco}},
  \bibinfo {author} {\bibfnamefont {Y.}~\bibnamefont {Huang}}, \bibinfo
  {author} {\bibfnamefont {V.}~\bibnamefont {Bisogni}}, \bibinfo {author}
  {\bibfnamefont {J.}~\bibnamefont {Pelliciari}}, \bibinfo {author}
  {\bibfnamefont {S.}~\bibnamefont {Fatale}}, \bibinfo {author} {\bibfnamefont
  {M.}~\bibnamefont {Dantz}}, \bibinfo {author} {\bibfnamefont {J.~G.}\
  \bibnamefont {Vale}}, \bibinfo {author} {\bibfnamefont {E.~C.}\ \bibnamefont
  {Hunter}}, \bibinfo {author} {\bibfnamefont {J.}~\bibnamefont {Chang}},
  \bibinfo {author} {\bibfnamefont {V.~N.}\ \bibnamefont {Strocov}}, \bibinfo
  {author} {\bibfnamefont {R.~S.}\ \bibnamefont {Perry}}, \bibinfo {author}
  {\bibfnamefont {M.}~\bibnamefont {Grioni}}, \bibinfo {author} {\bibfnamefont
  {D.~F.}\ \bibnamefont {McMorrow}}, \bibinfo {author} {\bibfnamefont {H.~M.}\
  \bibnamefont {R\o{}nnow}}, \ and\ \bibinfo {author} {\bibfnamefont
  {T.}~\bibnamefont {Schmitt}},\ }\href {\doibase 10.1103/PhysRevB.97.041102}
  {\bibfield  {journal} {\bibinfo  {journal} {Phys. Rev. B}\ }\textbf {\bibinfo
  {volume} {97}},\ \bibinfo {pages} {041102} (\bibinfo {year}
  {2018})}\BibitemShut {NoStop}%
\bibitem [{\citenamefont {Mizokawa}\ \emph {et~al.}(2001)\citenamefont
  {Mizokawa}, \citenamefont {Tjeng}, \citenamefont {Sawatzky}, \citenamefont
  {Ghiringhelli}, \citenamefont {Tjernberg}, \citenamefont {Brookes},
  \citenamefont {Fukazawa}, \citenamefont {Nakatsuji},\ and\ \citenamefont
  {Maeno}}]{Mizokawa:01}%
  \BibitemOpen
  \bibfield  {author} {\bibinfo {author} {\bibfnamefont {T.}~\bibnamefont
  {Mizokawa}}, \bibinfo {author} {\bibfnamefont {L.~H.}\ \bibnamefont {Tjeng}},
  \bibinfo {author} {\bibfnamefont {G.~A.}\ \bibnamefont {Sawatzky}}, \bibinfo
  {author} {\bibfnamefont {G.}~\bibnamefont {Ghiringhelli}}, \bibinfo {author}
  {\bibfnamefont {O.}~\bibnamefont {Tjernberg}}, \bibinfo {author}
  {\bibfnamefont {N.~B.}\ \bibnamefont {Brookes}}, \bibinfo {author}
  {\bibfnamefont {H.}~\bibnamefont {Fukazawa}}, \bibinfo {author}
  {\bibfnamefont {S.}~\bibnamefont {Nakatsuji}}, \ and\ \bibinfo {author}
  {\bibfnamefont {Y.}~\bibnamefont {Maeno}},\ }\href {\doibase
  10.1103/PhysRevLett.87.077202} {\bibfield  {journal} {\bibinfo  {journal}
  {Phys. Rev. Lett.}\ }\textbf {\bibinfo {volume} {87}},\ \bibinfo {pages}
  {077202} (\bibinfo {year} {2001})}\BibitemShut {NoStop}%
\bibitem [{\citenamefont {Thole}\ \emph {et~al.}(1992)\citenamefont {Thole},
  \citenamefont {Carra}, \citenamefont {Sette},\ and\ \citenamefont {van~der
  Laan}}]{Thole:92}%
  \BibitemOpen
  \bibfield  {author} {\bibinfo {author} {\bibfnamefont {B.~T.}\ \bibnamefont
  {Thole}}, \bibinfo {author} {\bibfnamefont {P.}~\bibnamefont {Carra}},
  \bibinfo {author} {\bibfnamefont {F.}~\bibnamefont {Sette}}, \ and\ \bibinfo
  {author} {\bibfnamefont {G.}~\bibnamefont {van~der Laan}},\ }\href {\doibase
  10.1103/PhysRevLett.68.1943} {\bibfield  {journal} {\bibinfo  {journal}
  {Phys. Rev. Lett.}\ }\textbf {\bibinfo {volume} {68}},\ \bibinfo {pages}
  {1943} (\bibinfo {year} {1992})}\BibitemShut {NoStop}%
\bibitem [{\citenamefont {Carra}\ \emph {et~al.}(1993)\citenamefont {Carra},
  \citenamefont {Thole}, \citenamefont {Altarelli},\ and\ \citenamefont
  {Wang}}]{Carra:93}%
  \BibitemOpen
  \bibfield  {author} {\bibinfo {author} {\bibfnamefont {P.}~\bibnamefont
  {Carra}}, \bibinfo {author} {\bibfnamefont {B.~T.}\ \bibnamefont {Thole}},
  \bibinfo {author} {\bibfnamefont {M.}~\bibnamefont {Altarelli}}, \ and\
  \bibinfo {author} {\bibfnamefont {X.}~\bibnamefont {Wang}},\ }\href {\doibase
  10.1103/PhysRevLett.70.694} {\bibfield  {journal} {\bibinfo  {journal} {Phys.
  Rev. Lett.}\ }\textbf {\bibinfo {volume} {70}},\ \bibinfo {pages} {694}
  (\bibinfo {year} {1993})}\BibitemShut {NoStop}%
\bibitem [{\citenamefont {Bun{\u a}u}\ and\ \citenamefont
  {Joly}(2009)}]{Bunau:09}%
  \BibitemOpen
  \bibfield  {author} {\bibinfo {author} {\bibfnamefont {O.}~\bibnamefont
  {Bun{\u a}u}}\ and\ \bibinfo {author} {\bibfnamefont {Y.}~\bibnamefont
  {Joly}},\ }\href {\doibase 10.1088/0953-8984/21/34/345501} {\bibfield
  {journal} {\bibinfo  {journal} {J. Phys.: Condens. Matter}\ }\textbf
  {\bibinfo {volume} {21}},\ \bibinfo {pages} {345501} (\bibinfo {year}
  {2009})}\BibitemShut {NoStop}%
\bibitem [{\citenamefont {Guda}\ \emph {et~al.}(2015)\citenamefont {Guda},
  \citenamefont {Guda}, \citenamefont {Soldatov}, \citenamefont {Lomachenko},
  \citenamefont {Bugaev}, \citenamefont {Lamberti}, \citenamefont {Gawelda},
  \citenamefont {Bressler}, \citenamefont {Smolentsev}, \citenamefont
  {Soldatov},\ and\ \citenamefont {Joly}}]{Guda:15}%
  \BibitemOpen
  \bibfield  {author} {\bibinfo {author} {\bibfnamefont {S.~A.}\ \bibnamefont
  {Guda}}, \bibinfo {author} {\bibfnamefont {A.~A.}\ \bibnamefont {Guda}},
  \bibinfo {author} {\bibfnamefont {M.~A.}\ \bibnamefont {Soldatov}}, \bibinfo
  {author} {\bibfnamefont {K.~A.}\ \bibnamefont {Lomachenko}}, \bibinfo
  {author} {\bibfnamefont {A.~L.}\ \bibnamefont {Bugaev}}, \bibinfo {author}
  {\bibfnamefont {C.}~\bibnamefont {Lamberti}}, \bibinfo {author}
  {\bibfnamefont {W.}~\bibnamefont {Gawelda}}, \bibinfo {author} {\bibfnamefont
  {C.}~\bibnamefont {Bressler}}, \bibinfo {author} {\bibfnamefont
  {G.}~\bibnamefont {Smolentsev}}, \bibinfo {author} {\bibfnamefont {A.~V.}\
  \bibnamefont {Soldatov}}, \ and\ \bibinfo {author} {\bibfnamefont
  {Y.}~\bibnamefont {Joly}},\ }\href@noop {} {\bibfield  {journal} {\bibinfo
  {journal} {J. Chem. Theory Comput.}\ }\textbf {\bibinfo {volume} {11}},\
  \bibinfo {pages} {4512} (\bibinfo {year} {2015})}\BibitemShut {NoStop}%
\bibitem [{SM()}]{SM}%
  \BibitemOpen
  \href@noop {} {}\bibinfo {howpublished} {See Supplemental Material at [URL
  will be inserted by publisher] for details on the calculated XAS spectra
  using the FDMNES code and on the inverse DM mechanism envisaged for the
  origin of anomalous dielectric loss at lower temperatures.}\BibitemShut
  {Stop}%
\bibitem [{\citenamefont {Kenzelmann}\ \emph {et~al.}(2005)\citenamefont
  {Kenzelmann}, \citenamefont {Harris}, \citenamefont {Jonas}, \citenamefont
  {Broholm}, \citenamefont {Schefer}, \citenamefont {Kim}, \citenamefont
  {Zhang}, \citenamefont {Cheong}, \citenamefont {Vajk},\ and\ \citenamefont
  {Lynn}}]{Kenzelmann:05}%
  \BibitemOpen
  \bibfield  {author} {\bibinfo {author} {\bibfnamefont {M.}~\bibnamefont
  {Kenzelmann}}, \bibinfo {author} {\bibfnamefont {A.~B.}\ \bibnamefont
  {Harris}}, \bibinfo {author} {\bibfnamefont {S.}~\bibnamefont {Jonas}},
  \bibinfo {author} {\bibfnamefont {C.}~\bibnamefont {Broholm}}, \bibinfo
  {author} {\bibfnamefont {J.}~\bibnamefont {Schefer}}, \bibinfo {author}
  {\bibfnamefont {S.~B.}\ \bibnamefont {Kim}}, \bibinfo {author} {\bibfnamefont
  {C.~L.}\ \bibnamefont {Zhang}}, \bibinfo {author} {\bibfnamefont {S.-W.}\
  \bibnamefont {Cheong}}, \bibinfo {author} {\bibfnamefont {O.~P.}\
  \bibnamefont {Vajk}}, \ and\ \bibinfo {author} {\bibfnamefont {J.~W.}\
  \bibnamefont {Lynn}},\ }\href {\doibase 10.1103/PhysRevLett.95.087206}
  {\bibfield  {journal} {\bibinfo  {journal} {Phys. Rev. Lett.}\ }\textbf
  {\bibinfo {volume} {95}},\ \bibinfo {pages} {087206} (\bibinfo {year}
  {2005})}\BibitemShut {NoStop}%
\bibitem [{\citenamefont {Yamasaki}\ \emph {et~al.}(2007)\citenamefont
  {Yamasaki}, \citenamefont {Sagayama}, \citenamefont {Goto}, \citenamefont
  {Matsuura}, \citenamefont {Hirota}, \citenamefont {Arima},\ and\
  \citenamefont {Tokura}}]{Yamazaki:07}%
  \BibitemOpen
  \bibfield  {author} {\bibinfo {author} {\bibfnamefont {Y.}~\bibnamefont
  {Yamasaki}}, \bibinfo {author} {\bibfnamefont {H.}~\bibnamefont {Sagayama}},
  \bibinfo {author} {\bibfnamefont {T.}~\bibnamefont {Goto}}, \bibinfo {author}
  {\bibfnamefont {M.}~\bibnamefont {Matsuura}}, \bibinfo {author}
  {\bibfnamefont {K.}~\bibnamefont {Hirota}}, \bibinfo {author} {\bibfnamefont
  {T.}~\bibnamefont {Arima}}, \ and\ \bibinfo {author} {\bibfnamefont
  {Y.}~\bibnamefont {Tokura}},\ }\href {\doibase 10.1103/PhysRevLett.98.147204}
  {\bibfield  {journal} {\bibinfo  {journal} {Phys. Rev. Lett.}\ }\textbf
  {\bibinfo {volume} {98}},\ \bibinfo {pages} {147204} (\bibinfo {year}
  {2007})}\BibitemShut {NoStop}%
\bibitem [{\citenamefont {Bogdanov}\ \emph {et~al.}(2015)\citenamefont
  {Bogdanov}, \citenamefont {Katukuri}, \citenamefont {Romh{\'a}nyi},
  \citenamefont {Yushankhai}, \citenamefont {Kataev}, \citenamefont
  {B{\"u}chner}, \citenamefont {van~den Brink},\ and\ \citenamefont
  {Hozoi}}]{Bogdanov:15}%
  \BibitemOpen
  \bibfield  {author} {\bibinfo {author} {\bibfnamefont {N.~A.}\ \bibnamefont
  {Bogdanov}}, \bibinfo {author} {\bibfnamefont {V.~M.}\ \bibnamefont
  {Katukuri}}, \bibinfo {author} {\bibfnamefont {J.}~\bibnamefont
  {Romh{\'a}nyi}}, \bibinfo {author} {\bibfnamefont {V.}~\bibnamefont
  {Yushankhai}}, \bibinfo {author} {\bibfnamefont {V.}~\bibnamefont {Kataev}},
  \bibinfo {author} {\bibfnamefont {B.}~\bibnamefont {B{\"u}chner}}, \bibinfo
  {author} {\bibfnamefont {J.}~\bibnamefont {van~den Brink}}, \ and\ \bibinfo
  {author} {\bibfnamefont {L.}~\bibnamefont {Hozoi}},\ }\href {\doibase
  10.1038/ncomms8306} {\bibfield  {journal} {\bibinfo  {journal} {Nature
  Commun.}\ }\textbf {\bibinfo {volume} {6}},\ \bibinfo {pages} {7306}
  (\bibinfo {year} {2015})}\BibitemShut {NoStop}%
\end{thebibliography}
%

\newpage
\setcounter{section}{0}
\setcounter{figure}{0}
\setcounter{table}{0}
\setcounter{equation}{0}
\renewcommand{\thefigure}{S\arabic{figure}}
\renewcommand{\thetable}{S\arabic{table}}
\renewcommand{\theequation}{S\arabic{equation}}

\noindent
\textbf{Supplemental Material: Direct observation of oxygen polarization in Sr$_2$IrO$_4$ by O $K$-edge x-ray magnetic circular dichroism}
\newline 
\begin{center}
{R. Kadono {\it et al.}}
\end{center}

\section{O $K$-edge XAS spectra calculated by the FDMNES code}
\begin{figure}[b]
  \centering
  \includegraphics[width=0.35\textwidth]{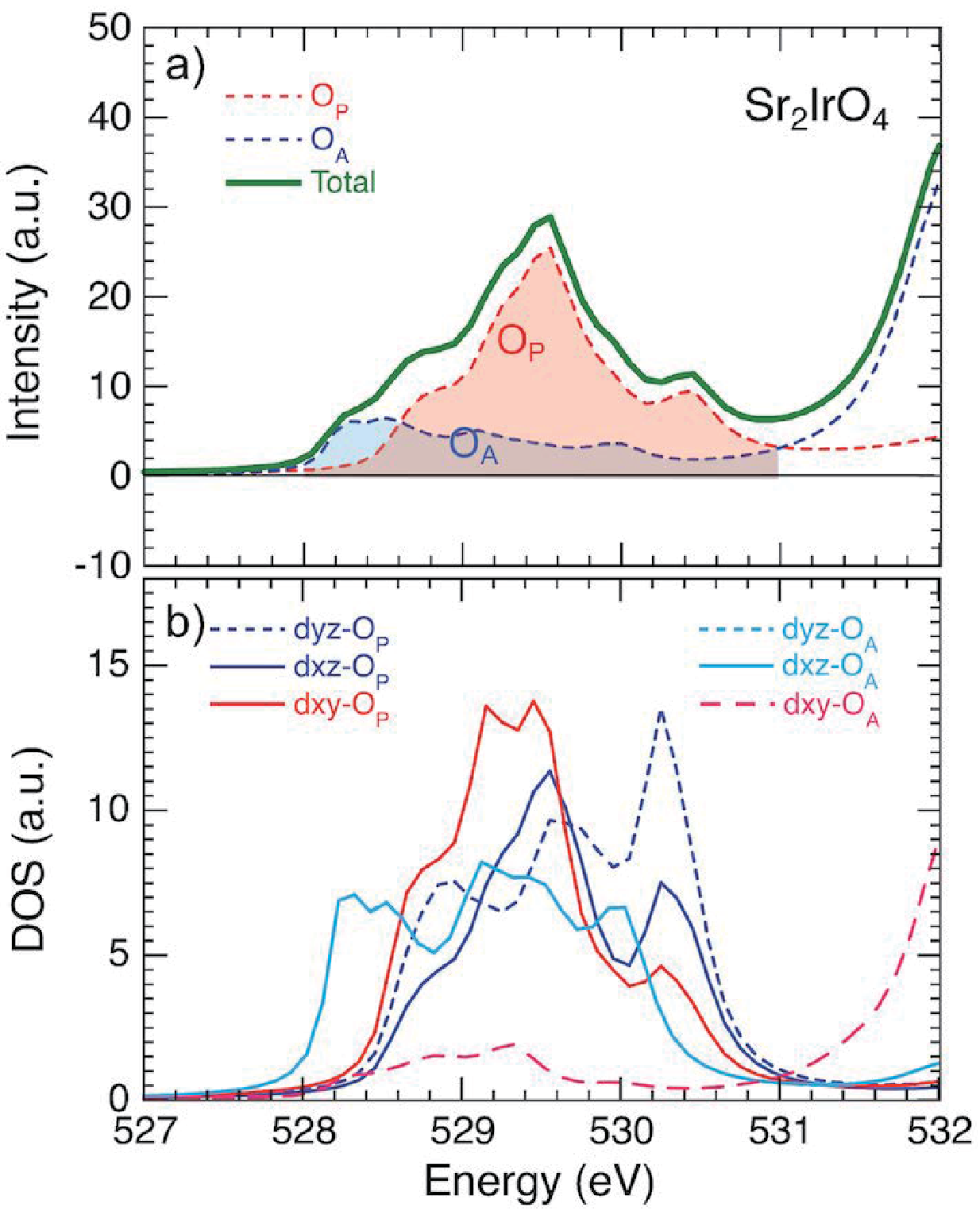}
  \vspace{-2mm}
  \caption{(a) XAS and (b) partial density of states (DOS) spectra for circularly polarized incident X-rays calculated using the FDMNES code, with ${\bm k}$-vector set to (0 0.8192 0.5736) corresponding to 55$^\circ$ in the $bc$ plane. O$_{\rm P}$ and O$_{\rm A}$ denote the planar and apical oxygen.  The hatched areas in (a) represent the regions integrated to evaluate the O$_{\rm A}$ and O$_{\rm P}$ contributions, respectively. The curves with $dxy$, $dyz$, and $dxz$ in (b) represent the contribution of Ir $5d$ $xy$, $yz$, and $xz$ orbitals.}
  \label{FigS1}
\end{figure}

We have conducted theoretical calculations on the O $K$-edge XAS spectra under circular polarization observed in \sio\ to evaluate the contributions from planar (O$_{\rm P}$) and apical  (O$_{\rm A}$) oxygen using the FDMNES code \cite{Bunau:09,Guda:15}. The calculations were performed under a variety of input conditions, and the result best reproduce the energy dependence of the observed XAS spectra is shown in Fig~\ref{FigS1}(a). In this calculation, we used the Muffin-tin approximation with a cluster radius of 7 \AA, and considered spin-orbit interaction and the on-site Coulomb energy (Hubbard $U = 2$ eV) for Ir. We did not consider the screening due to oxygen excitation. Instead, the appropriate energy shift was assumed in the convolution calculation (to produce a Lorentzian absorption spectrum with the contribution from the occupied state removed) to reproduce the pre-edge spectra. The same convolution procedure was applied to the calculated partial density of states (DOS) for O$_{\rm P}$ and O$_{\rm A}$, where the obtained DOS spectra for Ir 5$d$ $t_{2g}$ orbitals are shown in Fig~\ref{FigS1}(b).

The calculated result (solid line) in Fig~\ref{FigS1}(a) shows that the peak structures around 528.7 eV and 529.4 eV in the observed XAS spectra are well reproduced, although the linewidth is slightly wider and the shape is more complicated. It is also noticeable that the peak due to the hybridization between the O$_{\rm A}$ and $e_g$ orbitals appears at higher energies than the experimental value.  The corresponding partial DOS spectra in Fig~\ref{FigS1}(b) indicate that the XAS intensity over the energy range below 531 eV (hatched areas in Fig.~\ref{FigS1}) is dominated by the O 2$p$-Ir 5$d$ $t_{2g}$ hybridization, and the relative ratio of the O$_{\rm A}$ and O$_{\rm P}$ signal intensities over the corresponding energy region is calculated to be O$_{\rm A}$ : O$_{\rm P}$ = 1: 2.71. Although the separation of the O$_{\rm A}$ and O$_{\rm P}$ spectra is not good in this calculation, the obtained ratio agrees with the experimental value [O$_{\rm A}$ : O$_{\rm P}$ = 1 : 2.39(24)] within the  error, supporting our interpretation. 

Furthermore, looking at the partial DOS of the Ir $t_{2g}$ orbitals contributing to the XAS spectra of O$_{\rm P}$ and O$_{\rm A}$ in the same energy region, it can be seen that all $xy$, $yz$, and $xz$ orbitals contribute almost equally in the O$_{\rm P}$ XAS intensity, whereas there is almost no contribution from $xy$ in that of O$_{\rm A}$. The ratio of the contribution of the $xy$ and $yz+zx$ orbitals to the O$_{\rm P}$ XAS intensity is calculated to be $1:1.90$.

\section{Possible anti-ferroelectric correlation due to the spin current mechanism}
Cycloidal helical magnetism is one of the most typical causes of multiferroic behavior accompanying ferroelectricity. In the most general case, when two angled magnetic moments ${\bm S}_i$ and ${\bm S}_j$ with spin-orbit interactions are aligned across an anion, the electron distribution is expected to shift in the direction perpendicular to the vector connecting via the anions \cite{Katsura:05}. This phenomenon can be expressed by the following equation,
\begin{equation}
{\bm p}=\Lambda{\bm V}_{ij}\times({\bm S}_i\times{\bm S}_j),\label{SC}
\end{equation}
where ${\bm p}$ is the induced electric dipole moment, $\Lambda$ is a constant that depends on the spin-orbit interaction and the hopping integral, and ${\bm V}_{ij}$ is the vector connecting the two spin sites. This expression is closely related to the Hamiltonian for the Dzyaloshinsky-Moriya (DM) interaction,
\begin{equation}
{\mathcal H}=D(\theta)\cdot({\bm S}_i\times{\bm S}_j),\label{DM}
\end{equation}
which explains the phenomenon that the two spins tilt when the coupling angle $\theta$ determined by the crystal structure is finite. However, when the spins are fixed in Eq.~(\ref{DM}) and $\theta$ is taken as a variable, it has the same meaning as in Eq.~(\ref{SC}). This effect of a tilted spin pair giving rise to an electric dipole moment is called the inverse DM interaction. Although Eq.~(\ref{SC}) originally focused on the electron distribution and ignored the atomic displacement \cite{Katsura:05}, it holds for both cases. 

Now, when $({\bm S}_i\times{\bm S}_j)$ is always in the same direction, the electric dipole moments will also add up to produce a macroscopic electric polarization. Considering a one-dimensional spin chain, this condition is satisfied for the cycloidal helix where magnetic modulation vector and the spin rotation plane are parallel. For a common modulation vector,  the electric polarization is reversed when the direction of spin rotation is opposite. Such an electric polarization generation mechanism is also called a ``spin current mechanism'' \cite{Sergienko:06},  and it has been experimentally verified that many materials, including TbMnO$_3$ \cite{Kenzelmann:05,Yamazaki:07}, exhibit multiferroic ordering by this mechanism.
\begin{figure}[b]
  \centering
  \includegraphics[width=0.45\textwidth]{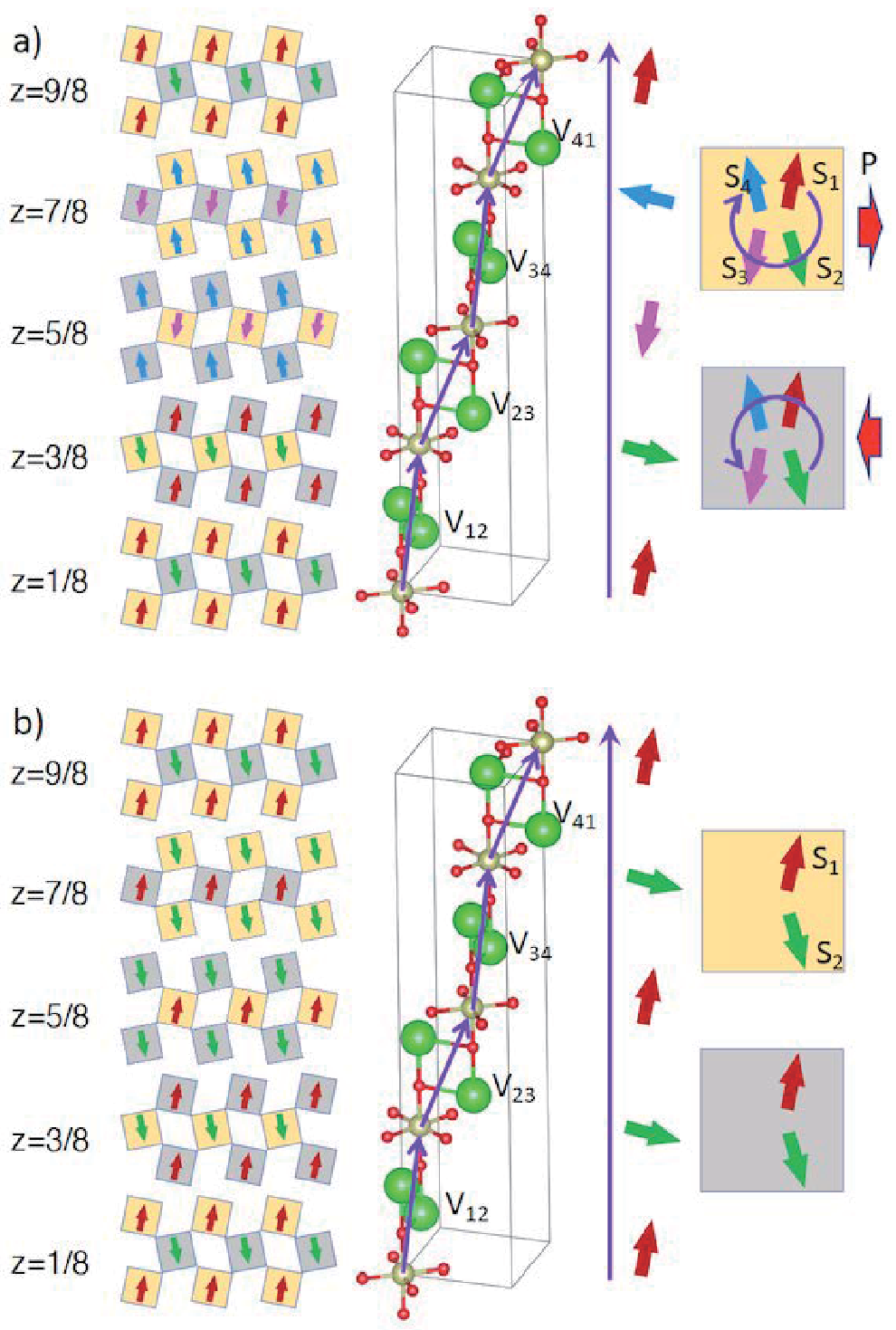}
  \vspace{-2mm}
  \caption{Variation along the $c$-axis of the Ir spin configuration in the $ab$ plane in the NC-AFM phase (a) and the WFM phase (b).}
  \label{FigS2}
\end{figure}

As shown in Fig.~\ref{FigS2}(a), a closer look at the 3D magnetic structure of Ir in the NC-AFM phase of \sio\ shows that a series of the connection vectors ${\bm V}_{ij}$ can be taken along the $c$ axis to have a cycloidal component parallel to the helical plane of spin rotation [i.e., ${\bm V}_{ij}\times({\bm S}_i\times{\bm S}_j)\neq0$]. Therefore, the spin current mechanism may cause electric polarization perpendicular to such a modulation vector. On the other hand, since the cycloid along the vector connecting neighboring Ir sites rotates in the opposite direction, the electric polarization is also reversed and no bulk electric polarization is generated. The existence of such anti-ferroelectric correlations explains the experimental fact of the anomalous increase in ac-dielectric loss below $T_{\rm M}\sim100$ K \cite{Chikara:09}. The same consideration for the WFM phase in Fig.~\ref{FigS2}(b) shows that the magnetic structure along the same modulation vector as in Fig.~\ref{FigS2}(a) is ferrimagnetic and no microscopic electric polarization occurs.

The electric polarization induced by the spin current mechanism is expected to be accompanied by a change in the relative electron distribution (i.e., orbital hybridization) between cations and anions; given that the increase in anti-ferroelectric correlation below $T_{\rm M}$ indicates a phase transition (or crossover) to the final 3D spin structure of Ir, this would be accompanied by a change in the Ir-O Sr-O-Ir electron distribution along the magnetic modulation vector. The ordering of the magnetic moment of the apical oxygen suggested by $\mu$SR \cite{Miyazaki:15} can be interpreted as a consequence of modulated hybridization. It is interesting to note that an anomalous electron distribution involving Sr is suggested by electron spin resonance (ESR) experiments in the WFM phase at low temperatures (4 K) \cite{Bogdanov:15}.

\end{document}